\documentclass[trackchanges,twocolumn]{aastex701}

\usepackage{booktabs}
\usepackage{multirow}
\usepackage{supertabular}
\usepackage{longtable}

\begin{document}

\title{From Variability to SED Modeling: A Multiwavelength Study of the Neutrino Blazar TXS~0506+056}


\author[orcid=0009-0004-0627-7139]{ShiYu Du} 
\affiliation{School of Physics and Astronomy, Beijing Normal University, Beijing 100875, China}
\affiliation{Institute for Frontiers in Astronomy and Astrophysics, Beijing Normal University, Beijing 102206, China}
\email{202421101125@mail.bnu.edu.cn}

\author[orcid=0009-0007-1759-1803]{Hanxiao Xia}
\affiliation{School of Physics and Astronomy, Beijing Normal University, Beijing 100875, China}
\affiliation{Institute for Frontiers in Astronomy and Astrophysics, Beijing Normal University, Beijing 102206, China}
\email{202321160018@mail.bnu.edu.cn}

\correspondingauthor{Jianghua Wu}
\author[orcid=0000-0002-8709-6759]{Jianghua Wu$^\dagger$}
\affiliation{School of Physics and Astronomy, Beijing Normal University, Beijing 100875, China}
\affiliation{Institute for Frontiers in Astronomy and Astrophysics, Beijing Normal University, Beijing 102206, China}
\email[show]{jhwu@bnu.edu.cn}

\author[orcid=0000-0003-1878-9428]{Yue Fang}
\affiliation{College of Intelligent Systems Science and Engineering, Hubei Minzu University, Enshi 445000, China}
\email{fangyue@hbmzu.edu.cn}


\begin{abstract}

The blazar TXS~0506+056 is the first source that was reported to be associated with high-energy extragalactic neutrino events and is one of the major targets for multi-messenger studies. We carried out multi-wavelength optical monitoring of this object on 24 nights in the period from 2018 to 2023. The overall light curves exhibit a dimming trend superposed by some small-amplitude fluctuations, and intraday variability was detected on four nights. Bluer-when-brighter behaviors were observed on both intraday and long timescales and were more pronounced on long timescales, while a weak redder-when-brighter trend was detected on one night. No significant time lags were found between variations at different optical wavelengths. We also retrieved the multi-broadband data from some monitoring programs. The data reveal complex, asynchronous flaring in different wavebands. A cross-correlation analysis shows that the high-energy emission (optical to $\gamma$-ray) is co-spatial and leads the radio emission by a substantial time of $\sim$800--900 days, suggesting that the radio emission originates from a downstream region of the jet. We performed time-dependent lepto-hadronic modeling of the spectral energy distributions for three representative epochs—the 2017 neutrino-associated flare, a post-flare phase, and a deep quiescent state—revealing an evolution in the radiative properties of the emission regions. The modeling results provide a phenomenological framework for interpreting the long-term multiwavelength behavior of TXS~0506+056 in a multi-messenger context.

\end{abstract}


\section{Introduction} \label{sect:intro}

Blazars are a class of active galactic nuclei (AGNs) whose powerful relativistic jets are closely aligned with our line of sight \citep[e.g.,][]{1995PASP..107..803U}. Their emission is dominated by non-thermal radiation, which forms a characteristic double-humped structure in their spectral energy distributions (SEDs) \citep[e.g.,][]{1998MNRAS.299..433F}. The low-energy hump is attributed to synchrotron radiation from relativistic electrons, while the high-energy component is thought to originate from either inverse Compton scattering in leptonic models \citep[e.g.,][]{2007Ap&SS.307...69B} or processes involving relativistic protons in hadronic models \citep[e.g.][]{2003APh....18..593M}. 

 Blazars are generally divided into two subclasses, namely BL Lacertae objects (BL Lacs) and flat-spectrum radio quasars (FSRQs), primarily distinguished by the strength of their optical emission lines and other spectral characteristics \citep[e.g.,][]{1996MNRAS.281..425M,1997A&A...327...61G}. A hallmark of blazars is their strong flux variability across all electromagnetic wavelengths and all observable timescales, from minutes to years \citep{1995ARA&A..33..163W,2004A&A...422..505G}. The color or spectral changes are usually associated with the flux variability. So the color-magnitude diagrams (CMDs) are widely employed to investigate the spectral evolution and reveal bluer-when-brighter (BWB), redder-when-brighter (RWB), or achromatic behaviors \citep[e.g.,][]{1997A&A...327...61G,2005AJ....129.1818W,2009ApJ...694..174B,2012ApJ...756...13B}. Different subclasses of blazars may exhibit distinct color behaviors. In particular, BL Lac objects often display a BWB trend, whereas FSRQs more frequently show a RWB behavior\citep[e.g.,][]{2006A&A...450...39G,2012MNRAS.425.3002G,2017ApJ...844..107I}, although more complex patterns may arise depending on the observing passbands and the contribution of relatively stable emission components \citep[see][and references therein]{2011MNRAS.418.1640W}. The BWB trend frequently observed in BL Lac objects is generally interpreted as a manifestation of particle acceleration and cooling processes in the jet or variations of the Doppler factor acting on a curved synchrotron spectrum \citep{1985ApJ...298..114M,1998A&A...333..452K,2015MNRAS.454..353R,2021APh...12902577B}.

The blazar TXS~0506+056 is suggested to be a masquerading BL Lac or an FSRQ with hidden broad lines \citep{2019MNRAS.484L.104P}. It is at a redshift of $z=0.3365$ \citep{2018ApJ...854L..32P} and is the first reported extragalactic source to be associated with a high-energy neutrino event IC-170922A at a significance greater than $3\sigma$ \citep{2018Sci...361..147I}. An archival analysis subsequently revealed a neutrino outburst of $13 \pm 5$ events from the same location during a 158-day period in 2014--2015 \citep{2018Sci...361.1378I} and reinforced its importance as a likely site of cosmic-ray acceleration and neutrino production \citep{1995APh.....3..295M,1996SSRv...75..341S}.

Great efforts have been made to understand the multimessenger emission from TXS 0506+056. Rapid multiwavelength follow-up observations of this object after the detection of IC-170922A revealed pronounced variability from radio to $\gamma$-ray energies \citep[e.g.,][]{2018RNAAS...2..130G,2019A&A...630A.103B,2021PASJ...73...25M,2021BlgAJ..34...79B,2023mgm..conf.1467L}. \citet{2024MNRAS.527.1344D} carried out multiband optical monitoring of TXS~0506+056 and investigated its flux and color variability on diverse timescales. Nevertheless, systematic studies of optical intraday variability for this source remain relatively limited, highlighting the need for further dedicated monitoring. For the broadband emission of this object, a variety of leptonic, lepto-hadronic, and hadronic models have been proposed to reproduce its SED and the observed neutrino emissions \citep[e.g.,][]{2018ApJ...864...84K,2019NatAs...3...88G,2019ApJ...874L..29R}. While the standard one-zone models can generally describe the electromagnetic radiation, they often have difficulties in producing the required neutrino flux without conflicting with constraints from contemporaneous X-ray observations. To alleviate these tensions, more complex scenarios involving multiple radiation zones or additional target photon fields have been suggested \citep[e.g.,][]{2021ApJ...906...51X,2024ApJ...962..142W,2025PhRvD.112h3016W}. Despite these efforts, a unified model capable of simultaneously explaining both the electromagnetic emission and the observed neutrino flux has not yet been firmly established.

In this work, we performed dedicated multi-night intraday optical monitoring of this source, providing detailed studies of its optical variability on intraday and long timescales. Combining these observations with the retrieved long-term multi-broadband data from radio to $\gamma$-ray bands, we constructed its SEDs at three representative states and performed modeling to explore the underlying radiation mechanisms in the jet. The paper is organized as follows. Sections~\ref{sec:obs} describe the data collection and reduction. Section~\ref{sec:opt-var} presents the optical variability analysis based on our observations. In Section~\ref{sect:analysis}, we detail the long-term multi-broadband analysis and the SED modeling. Finally, the conclusions are given in Section~\ref{sect:conclusion}.

\section{Observations and Data Analysis} 
\label{sec:obs}
\subsection{Optical Observation and Data Reduction}

The intraday optical monitoring was conducted using the 85\,cm telescope at Xinglong Station of the National Astronomical Observatories of the Chinese Academy of Sciences (NAOC). The telescope uses a prime-focus optical design with a focal ratio of $F/3.3$. The CCD is a $2048 \times 2048$ chip with a field of view of  $\sim 32.8 \times 32.8~\mathrm{arcmin}^2$. We monitored TXS~0506+056 in the $R$ and $I$ bands in October, 2018 and in the $B$, $V$, and $R$ bands after that. About 3000 data points were collected on 24 nights between October 2018 and November 2023. The corresponding observation information is summarized in Table~\ref{Tab1}.

\begin{figure}[ht!]
\plotone{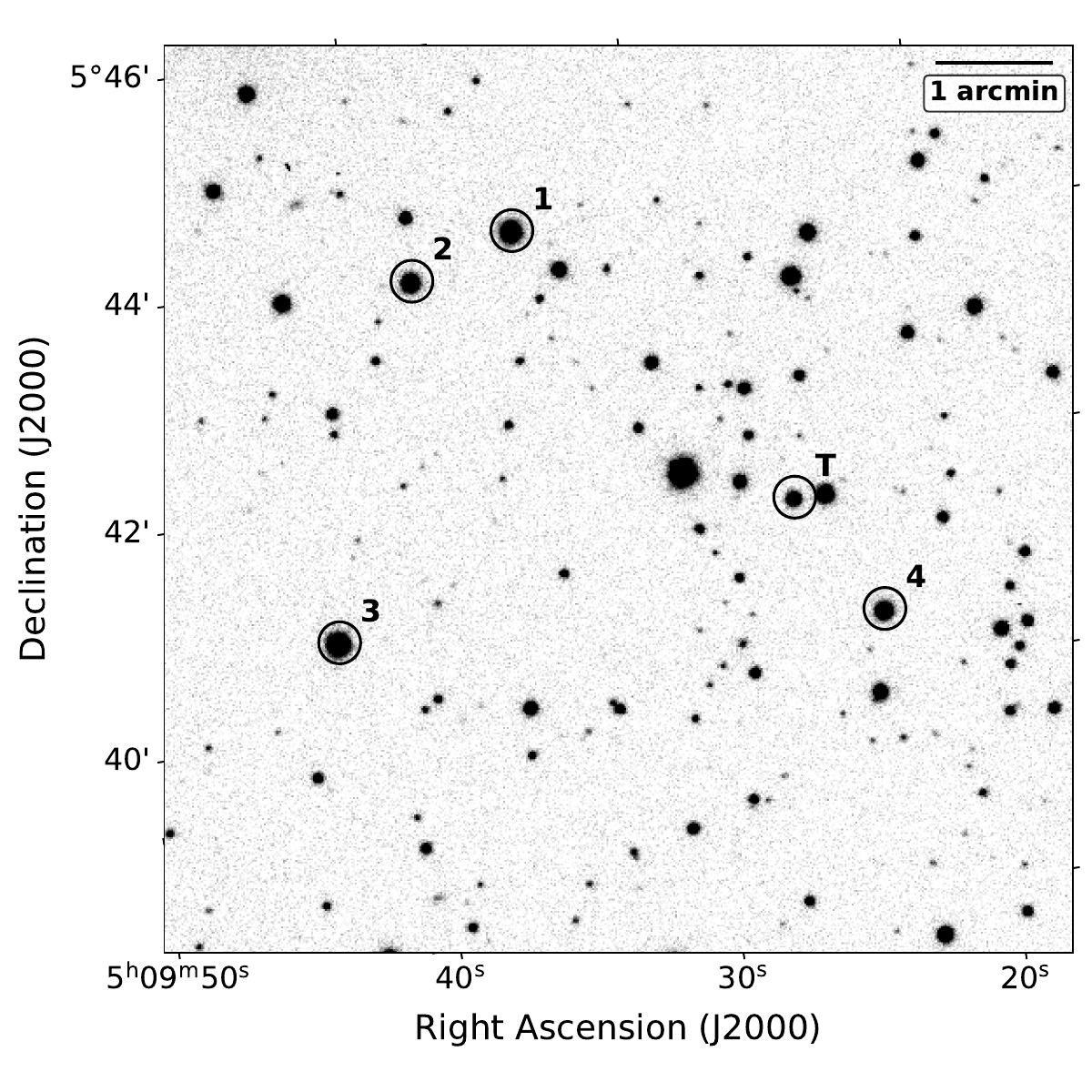}
\caption{Finding Chart of TXS~0506+056 Obtained in the $R$ Band on 2020 January 17. The target blazar, three comparison stars, and the check star are marked as ``T'' and ``1-4'', respectively.}
\label{Fig1}
\end{figure}

As shown in Figure~\ref{Fig1}, TXS~0506+056 and four stars are marked, where stars 1 , 2 and 3 are selected as comparison stars, and star 4 is used as the check star. The raw data were reduced with the standard procedure, including bias subtraction, flat-fielding, and aperture photometry. The data obtained before 2019 April were reduced using the Image Reduction and Analysis Facility (IRAF)\footnote{Distributed by the National Optical Astronomy Observatories (NOAO), operated by AURA under a cooperative agreement with the NSF.}, and the data afterwards were reduced with the \texttt{Photutils} package (version 1.12.0\footnote{\url{https://photutils.readthedocs.io/}}; \citealt{2024zndo..10967176B}). The photometry was performed using several aperture radii ranging from 1 to 3 times the FWHM to determine an optimal aperture size. Finally, an aperture radius of $2\times\mathrm{FWHM}$, which provided minimum standard deviations of the differential magnitudes between the check star and the comparison stars, was adopted. The fluxes of TXS~0506+056 were calibrated relative to there comparison stars. Some data points suffered from poor observing conditions or instrumental failures were excluded. The intraday light curves of TXS~0506+056 and the check star are presented in Figure~\ref{Fig2}, where the $B$- and $V$-band light curves are shifted vertically for clarity. Note that the error bars are slightly larger during the last three days, particularly at the beginning and end of MJD~60273 and the final third of MJD~60274, likely due to relatively poor weather conditions. The complete Xinglong 85 cm optical monitoring data are available in machine-readable format in the online journal.

\begin{deluxetable*}{cccccrrrrcrrcr}
\tabletypesize{\footnotesize} 
\tablecolumns{14} 
\tablewidth{0pt} 
\tablecaption{IDV Analysis Results of TXS~0506+056 \label{Tab1}}

\tablehead{
\colhead{MJD} & \colhead{Date} & \colhead{Filter} & \colhead{N} & \colhead{$\nu$} & 
\multicolumn{2}{c}{$\chi^2$ test} & \multicolumn{2}{c}{$F$-test} & 
\multicolumn{3}{c}{ANOVA test} & \colhead{Var?} & \colhead{Amp} \\
\cmidrule(lr){6-7} \cmidrule(lr){8-9} \cmidrule(lr){10-12}
\colhead{} & \colhead{} & \colhead{} & \colhead{} & \colhead{} & 
\colhead{$\chi^2$} & \colhead{$\chi^2_{\rm c}$} & \colhead{$F$} & \colhead{$F_{\rm c}$} & 
\colhead{$\nu_1,\nu_2$} & \colhead{$F$} & \colhead{$F_{\rm c}$} & \colhead{} & \colhead{(\%)}
}

\startdata
58417 & 20181026 & R & 14 & 13 & 359.55 & 27.69 & 27.18 & 3.91 & 2,11 & 2.79 & 7.21 & N &      \\
      &          & I & 14 & 13 & 89.36  & 27.69 & 4.42  & 3.91 & 2,11 & 1.67 & 7.21 & N &      \\
58419 & 20181028 & R & 16 & 15 & 81.20  & 30.58 & 4.66  & 3.52 & 3,12 & 5.74 & 5.95 & N &      \\
      &          & I & 14 & 13 & 20.75  & 27.69 & 0.14  & 3.91 & 2,11 & 2.94 & 7.21 & N &      \\
58420 & 20181029 & R & 16 & 15 & 949.28 & 30.58 & 95.64 & 3.52 & 3,12 & 12.49 & 5.95 & V & 9.48 \\
      &          & I & 16 & 15 & 1366.80 & 30.58 & 16.32 & 3.52 & 3,12 & 30.23 & 5.95 & V & 9.89 \\
58424 & 20181102 & B & 42 & 41 & 218.25 & 64.95 & 3.77 & 2.09 & 8,33 & 0.93 & 3.11 & N &      \\
      &          & V & 42 & 41 & 266.97 & 64.95 & 4.93 & 2.09 & 8,33 & 4.99 & 3.11 & V & 4.96 \\
      &          & R & 42 & 41 & 295.74 & 64.95 & 4.52 & 2.09 & 8,33 & 1.80 & 3.11 & N &      \\
58455 & 20181203 & B & 40 & 39 & 166.26 & 62.43 & 5.08 & 2.14 & 7,32 & 0.79 & 3.26 & N &      \\
      &          & V & 40 & 39 & 141.72 & 62.43 & 8.54 & 2.14 & 7,32 & 1.67 & 3.26 & N &      \\
      &          & R & 40 & 39 & 200.97 & 62.43 & 8.36 & 2.14 & 7,32 & 1.94 & 3.26 & N &      \\
58456 & 20181204 & B & 69 & 68 & 223.68 & 98.03 & 1.83 & 1.77 & 13,55 & 0.73 & 2.47 & N &      \\
      &          & V & 67 & 66 & 210.34 & 95.63 & 4.57 & 1.78 & 13,53 & 0.48 & 2.49 & N &      \\
      &          & R & 66 & 65 & 166.35 & 94.42 & 3.90 & 1.79 & 13,52 & 0.68 & 2.49 & N &      \\
58514 & 20190131 & B & 9 & 8 & 8.38 & 20.09 & 5.32 & 6.03 & 1,7 & 0.22 & 12.25 & N &      \\
      &          & V & 9 & 8 & 6.80 & 20.09 & 1.93 & 6.03 & 1,7 & 0.37 & 12.25 & N &      \\
      &          & R & 9 & 8 & 22.89 & 20.09 & 3.55 & 6.03 & 1,7 & 0.00 & 12.25 & N &      \\
58515 & 20190201 & B & 17 & 16 & 173.66 & 32.00 & 10.75 & 3.37 & 3,13 & 9.72 & 5.74 & V & 4.17 \\
      &          & V & 17 & 16 & 314.67 & 32.00 & 66.31 & 3.37 & 3,13 & 9.01 & 5.74 & V & 3.88 \\
      &          & R & 17 & 16 & 489.86 & 32.00 & 5.37 & 3.37 & 3,13 & 9.08 & 5.74 & V & 4.29 \\
58517 & 20190203 & B & 13 & 12 & 28.54 & 26.22 & 2.12 & 4.16 & 2,10 & 0.65 & 7.56 & N &      \\
      &          & V & 14 & 13 & 60.16 & 27.69 & 5.43 & 3.91 & 2,11 & 0.61 & 7.21 & N &      \\
      &          & R & 14 & 13 & 64.60 & 27.69 & 4.25 & 3.91 & 2,11 & 0.71 & 7.21 & N &      \\
58547 & 20190305 & B & 7 & 6 & 37.27 & 16.81 & 1.76 & 8.47 & 1,5 & 2.19 & 16.26 & N &      \\
      &          & V & 7 & 6 & 34.86 & 16.81 & 0.93 & 8.47 & 1,5 & 2.49 & 16.26 & N &      \\
      &          & R & 7 & 6 & 78.00 & 16.81 & 0.35 & 8.47 & 1,5 & 0.35 & 16.26 & N &      \\
58548 & 20190306 & B & 8 & 7 & 25.25 & 18.48 & 1.87 & 6.99 & 1,6 & 3.44 & 13.75 & N &      \\
      &          & V & 8 & 7 & 117.56 & 18.48 & 7.63 & 6.99 & 1,6 & 1.76 & 13.75 & N &      \\
      &          & R & 8 & 7 & 179.64 & 18.48 & 5.00 & 6.99 & 1,6 & 2.81 & 13.75 & N &      \\
58549 & 20190307 & B & 7 & 6 & 4.90 & 16.81 & 0.36 & 8.47 & 1,5 & 0.57 & 16.26 & N &      \\
      &          & V & 6 & 5 & 9.00 & 15.09 & 0.31 & 10.97 & 1,4 & 0.01 & 21.20 & N &      \\
      &          & R & 6 & 5 & 21.95 & 15.09 & 0.86 & 10.97 & 1,4 & 5.68 & 21.20 & N &      \\
58820 & 20191203 & B & 100 & 99  & 1041.48 & 134.64 & 3.96 & 1.60 & 19,80 & 4.59 & 2.14 & V & 11.36 \\
      &          & V & 100 & 99  & 390.33  & 134.64 & 3.90 & 1.60 & 19,80 & 1.79 & 2.14 & N &       \\
      &          & R & 100 & 99  & 329.73  & 134.64 & 4.00 & 1.60 & 19,80 & 1.04 & 2.14 & N &       \\
58821 & 20191204 & B & 60  & 59  & 455.14  & 87.17  & 15.76 & 1.85 & 11,48 & 4.18 & 2.64 & V & 9.51  \\
      &          & V & 60  & 59  & 448.83  & 87.17  & 20.54 & 1.85 & 11,48 & 4.81 & 2.64 & V & 6.82  \\
      &          & R & 59  & 58  & 550.17  & 85.95  & 17.07 & 1.86 & 11,47 & 6.14 & 2.65 & V & 7.72  \\
58822 & 20191205 & B & 60  & 59  & 83.70   & 87.17  & 3.72 & 1.85 & 11,48 & 0.58 & 2.64 & N &       \\
      &          & V & 60  & 59  & 82.65   & 87.17  & 4.48 & 1.85 & 11,48 & 0.35 & 2.64 & N &       \\
      &          & R & 60  & 59  & 97.89   & 87.17  & 3.82 & 1.85 & 11,48 & 0.38 & 2.64 & N &       \\
58823 & 20191206 & B & 60  & 59  & 86.54   & 87.17  & 6.07 & 1.85 & 11,48 & 0.33 & 2.64 & N &       \\
      &          & V & 60  & 59  & 85.23   & 87.17  & 4.64 & 1.85 & 11,48 & 0.23 & 2.64 & N &       \\
      &          & R & 60  & 59  & 90.39   & 87.17  & 5.44 & 1.85 & 11,48 & 0.31 & 2.64 & N &       \\
58861 & 20200113 & B & 21  & 20  & 19.53   & 37.57  & 3.29 & 2.94 & 4,16  & 0.15 & 4.77 & N &       \\
      &          & V & 21  & 20  & 39.73   & 37.57  & 5.15 & 2.94 & 4,16  & 0.55 & 4.77 & N &       \\
      &          & R & 21  & 20  & 24.96   & 37.57  & 2.06 & 2.94 & 4,16  & 0.60 & 4.77 & N &       \\
58862 & 20200114 & B & 55  & 54  & 182.93  & 81.07  & 6.31 & 1.90 & 10,44 & 2.50 & 2.75 & N &       \\
      &          & V & 55  & 54  & 130.04  & 81.07  & 4.56 & 1.90 & 10,44 & 1.06 & 2.75 & N &       \\
      &          & R & 55  & 54  & 92.37   & 81.07  & 3.38 & 1.90 & 10,44 & 1.34 & 2.75 & N &       \\
\enddata

\tablecomments{Continued on next page.}
\end{deluxetable*}

\begin{deluxetable*}{cccccrrrrcrrcr}
\tabletypesize{\footnotesize} 
\tablecolumns{14} 
\tablewidth{0pt} 
\tablecaption{IDV Analysis Results of TXS~0506+056 --- Continued}

\tablehead{
\colhead{MJD} & \colhead{Date} & \colhead{Filter} & \colhead{N} & \colhead{$\nu$} & 
\multicolumn{2}{c}{$\chi^2$ test} & \multicolumn{2}{c}{$F$-test} & 
\multicolumn{3}{c}{ANOVA test} & \colhead{Var?} & \colhead{Amp} \\
\cmidrule(lr){6-7} \cmidrule(lr){8-9} \cmidrule(lr){10-12}
\colhead{} & \colhead{} & \colhead{} & \colhead{} & \colhead{} & 
\colhead{$\chi^2$} & \colhead{$\chi^2_{\rm c}$} & \colhead{$F$} & \colhead{$F_{\rm c}$} & 
\colhead{$\nu_1,\nu_2$} & \colhead{$F$} & \colhead{$F_{\rm c}$} & \colhead{} & \colhead{(\%)}
}

\startdata
58863 & 20200115 & B & 73  & 72  & 105.28  & 102.82 & 3.03 & 1.74 & 14,58 & 0.36 & 2.41 & N &       \\
      &          & V & 72  & 71  & 104.68  & 101.62 & 2.44 & 1.75 & 14,57 & 0.27 & 2.41 & N &       \\
      &          & R & 74  & 73  & 131.53  & 104.01 & 2.22 & 1.73 & 14,59 & 0.13 & 2.40 & N &       \\
58865 & 20200117 & B & 42  & 41  & 327.80  & 64.95  & 19.72 & 2.09 & 8,33  & 5.83 & 3.11 & V & 9.31  \\
      &          & V & 42  & 41  & 256.04  & 64.95  & 15.03 & 2.09 & 8,33  & 1.98 & 3.11 & N &       \\
      &          & R & 42  & 41  & 525.43  & 64.95  & 4.76 & 2.09 & 8,33  & 6.44 & 3.11 & V & 7.36  \\
59159 & 20201106 & B & 27  & 26  & 50.19   & 45.64  & 0.99 & 2.55 & 5,21  & 3.78 & 4.04 & N &       \\
      &          & V & 27  & 26  & 23.82   & 45.64  & 0.77 & 2.55 & 5,21  & 0.46 & 4.04 & N &       \\
      &          & R & 27  & 26  & 249.69  & 45.64  & 1.11 & 2.55 & 5,21  & 0.72 & 4.04 & N &       \\
60273 & 20231125 & B & 130 & 129 & 169.38  & 169.28 & 7.56 & 1.51 & 25,104 & 0.43 & 1.96 & N &       \\
      &          & V & 130 & 129 & 199.29  & 169.28 & 9.40 & 1.51 & 25,104 & 0.55 & 1.96 & N &       \\
      &          & R & 130 & 129 & 245.44  & 169.28 & 8.81 & 1.51 & 25,104 & 0.41 & 1.96 & N &       \\
60274 & 20231126 & B & 105 & 104 & 149.07  & 140.46 & 7.88 & 1.58 & 20,84 & 0.51 & 2.10 & N &       \\
      &          & V & 105 & 104 & 123.22  & 140.46 & 9.18 & 1.58 & 20,84 & 0.60 & 2.10 & N &       \\
      &          & R & 105 & 104 & 172.25  & 140.46 & 12.70 & 1.58 & 20,84 & 1.16 & 2.10 & N &       \\
60275 & 20231127 & B & 93  & 92  & 411.99  & 126.46 & 11.84 & 1.63 & 18,74 & 0.51 & 2.19 & N &       \\
      &          & V & 93  & 92  & 118.42  & 126.46 & 10.61 & 1.63 & 18,74 & 0.85 & 2.19 & N &       \\
      &          & R & 93  & 92  & 996.97  & 126.46 & 33.05 & 1.63 & 18,74 & 9.64 & 2.19 & V & 29.90 \\
\enddata

\tablecomments{The columns are the Modified Julian Date, the calendar date, filter, the number of exposures, degree of freedom for $\chi^2$ test and $F$-test, $\chi^2 \backslash F$, and the critical value $\chi^2_{\rm c} \backslash F_{\rm c}$ at the 99$\%$ confidence level, two degrees of freedom, $F$, and $F_{\rm c}$ in the ANOVA test, variable or not (“V” is variable and “N” is nonvariable), and variability amplitude, respectively.
}
\end{deluxetable*}

\begin{figure*}[t!] 
\centering
\includegraphics[width=\textwidth]{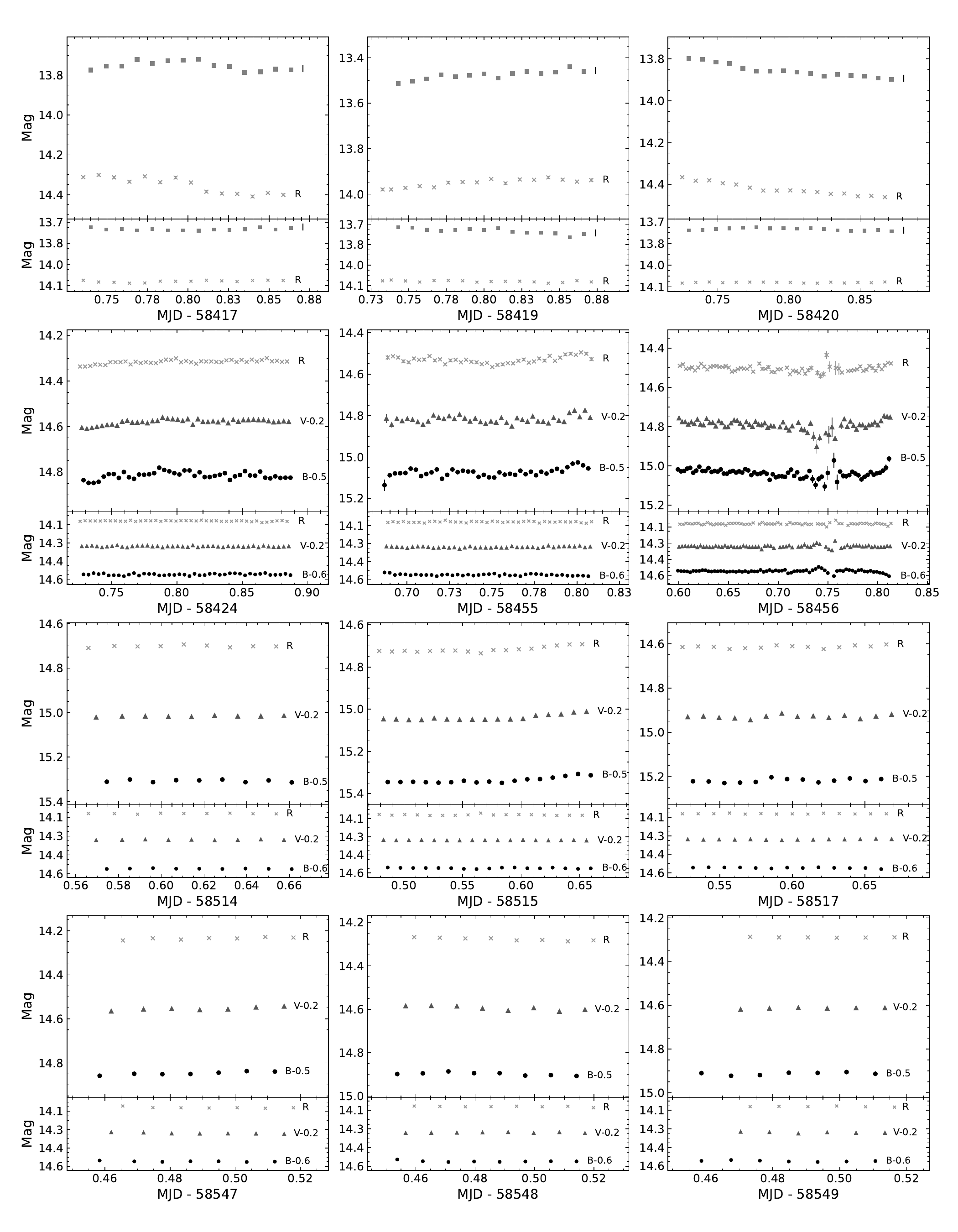}
\caption{Intraday Light Curves of TXS~0506+056 (large panels) and the Check Star (small panels). Only $R$ and $I$ bands are available for the first three days.}
\label{Fig2}
\end{figure*}

\begin{figure*}[t!]
\addtocounter{figure}{-1} 
\centering
\includegraphics[width=\textwidth]{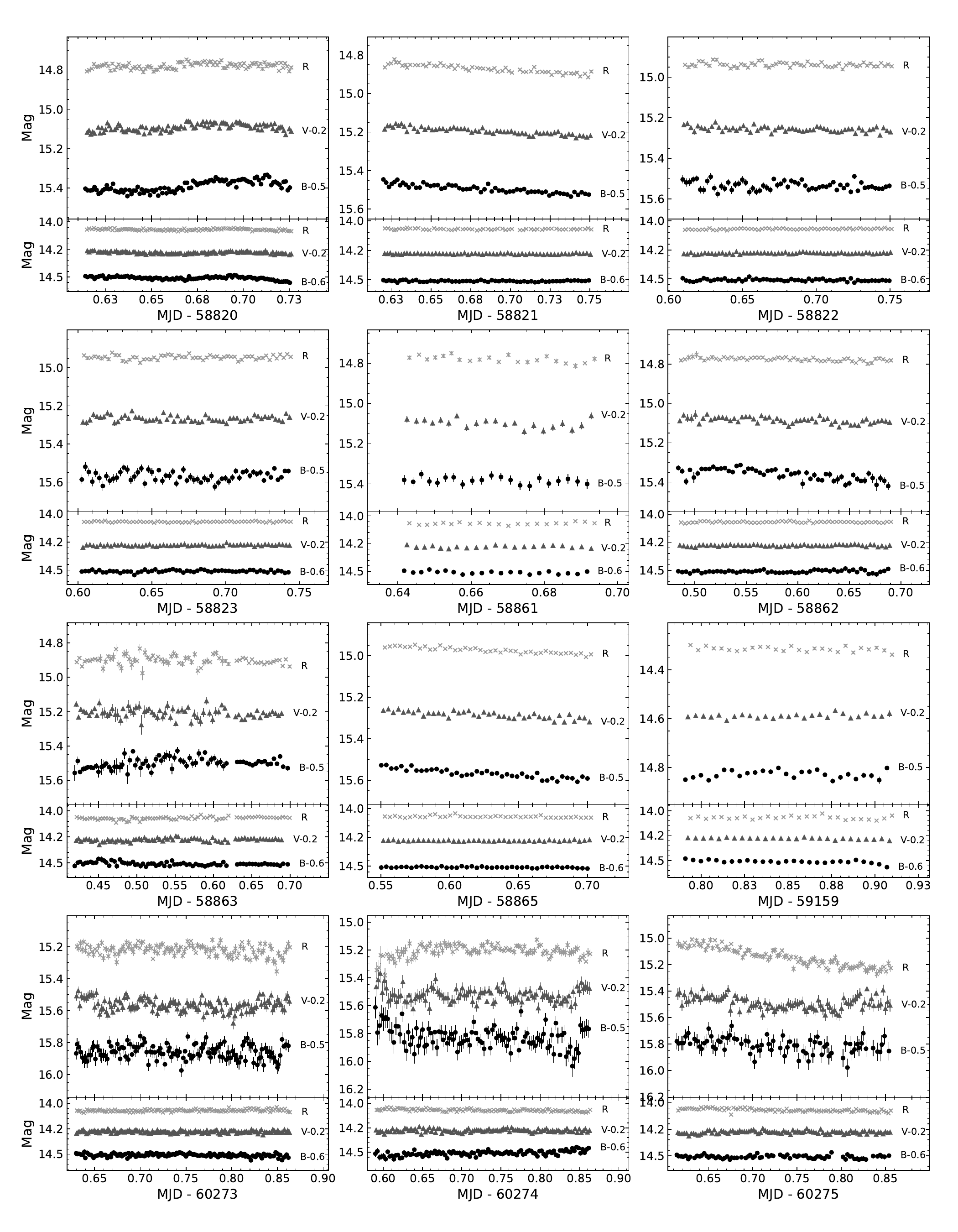}
\caption{Continued. Error bars are slightly larger for the last three days.}
\end{figure*}

\twocolumngrid 

\subsection{Multi-broadband Data Acquisition} \label{sec:data}
\subsubsection{Optical}

To extend the temporal coverage of the optical data for TXS~0506+056, we collected archival photometric data from several public databases, including those of the American Association of Variable Star Observers (AAVSO)\footnote{\url{https://www.aavso.org}}, the Zwicky Transient Facility (ZTF, \citealt{2019PASP..131a8002B,2019PASP..131a8003M})\footnote{\url{https://irsa.ipac.caltech.edu/Missions/ztf.html}}, the All-Sky Automated Survey for Supernovae (ASAS-SN, \citealt{2019MNRAS.485..961J,2014ApJ...788...48S,2023arXiv230403791H})\footnote{\url{http://asas-sn.ifa.hawaii.edu/skypatrol/}}, and the Katzman Automatic Imaging Telescope (KAIT)\footnote{\url{http://herculesii.astro.berkeley.edu/kait/agn/}}. The data have a time span from MJD~55800 to 60600 and include the photometric measurements in the $BVRI$ bands of AAVSO, the $gri$ bands of ZTF, the $Vg$ bands from ASAS-SN, and the $R$-band from KAIT. Due to the relatively sparse sampling, the $B$ and $I/i$ bands were not considered in the following analysis.

\subsubsection{\texorpdfstring{$\gamma$}{gamma}-ray}

The $0.1$ -- $100~\mathrm{GeV}$ $\gamma$-ray light curve of TXS~0506+056 from the Fermi-LAT Lightcurve Repository (LCR,\citealt{2023ApJS..265...31A})\footnote{\url{https://fermi.gsfc.nasa.gov/ssc/data/access/lat/ LightCurveRepository/}} was retrieved with a time bin of 3 days and a period from MJD~55000 to 60600.

For the SED fitting, Pass~8 Fermi-LAT data in the $0.1$--$300~\mathrm{GeV}$ energy range were obtained from the Fermi Science Support Center (FSSC)\footnote{\url{https://fermi.gsfc.nasa.gov/ssc/data/access/}}. An unbinned likelihood analysis was carried out with version~2.4.0 of the Fermi Science Tools (\citealt{2019ascl.soft05011F}). Events were selected from a $20^\circ$ region of interest (ROI) centered on TXS~0506+056 (RA~=~77.359$^\circ$, Dec~=~5.701$^\circ$) using SOURCE class events (\texttt{evclass~=~128}, \texttt{evtype~=~3}) and a zenith angle cut of \texttt{zmax~=~90}$^\circ$. The source model was constructed with the \texttt{make4FGLxml.py} script, including the standard Galactic diffuse emission (\texttt{gll\_iem\_v07.fits}) and isotropic background (\texttt{iso\_P8R3\_SOURCE\_V3\_v1.txt}). The spectral parameters of sources within $5^\circ$ of the ROI center were left free, while those outside this region were fixed to their 4FGL values \citep{2023arXiv230712546B}. The $\gamma$-ray spectrum of TXS~0506+056 was fitted with a log-parabola model in the likelihood analysis, from which the SED spectral points were derived, and detections with TS~$>$~10 were regarded as significant.

\subsubsection{X-ray}

The X-ray observations of TXS~0506+056 were obtained with the Swift X-ray Telescope (XRT; \citealt{2005SSRv..120..165B}), operating in the 0.3--10~keV energy range. The X-ray light curves and spectra were retrieved with the online Swift-XRT data products generator tool provided by the UK Swift Science Data Centre\citep{2009MNRAS.397.1177E}\footnote{\url{https://www.swift.ac.uk/user_objects/}} , and were processed using \textsc{HEASOFT} version 6.35.2.

For each observation, source events were extracted from a circular region with a radius of 1~arcmin centered on the source position. The resulting spectra were grouped with the \texttt{grppha} tool and analyzed using the Cash statistic \citep{1979ApJ...228..939C}. The spectral fitting was performed using \textsc{XSPEC} version 12.15.0. The spectra were modeled with an absorbed power-law, where photoelectric absorption was described by the \texttt{tbabs} model \citep{2000ApJ...542..914W}. Both Galactic and intrinsic absorption components were included. The Galactic hydrogen column density was fixed at $N_{\rm H} = 1.55 \times 10^{21}~\mathrm{cm^{-2}}$ \citep{2013MNRAS.431..394W}, while the intrinsic absorption was left free during the fitting. The best-fit intrinsic absorption was negligible for all observations.

\subsubsection{Radio}

The radio data of TXS~0506+056 were collected from several long-term monitoring programs at millimeter and centimeter wavelengths. The 1~mm band light curve was obtained from the Submillimeter Array (SMA; \citealt{2004ApJ...616L...1H}) calibrator list\footnote{\url{http://sma1.sma.hawaii.edu/callist/callist.html}}, which provides flux density measurements at 1~mm and 850~$\mu$m for a large number of compact radio sources. At centimeter wavelengths, the 37~GHz light curve was taken from the Metsähovi radio observatory monitoring program\footnote{\url{https://www.aalto.fi/en/metsahovi-radio-observatory}}. In addition, the 15~GHz data were adopted from the VLBA MOJAVE program\footnote{\url{https://www.cv.nrao.edu/MOJAVE/sourcepages/}} \citep{2018ApJS..234...12L}. These radio light curves were used in the subsequent multiwavelength analysis.

\section{Optical Variability Revealed by Our Observations } \label{sec:opt-var}
\subsection{Intraday Variability Detection}

To quantitatively assess the presence of intraday optical variability (IDV) in TXS~0506+056, we applied three commonly used statistical tests to the intraday light curves, namely the $\chi^{2}$  test (\citealt{2012MNRAS.420.3147G}), the $F$-test (\citealt{2010AJ....139.1269D}), and the one-way analysis of variance (ANOVA; \citealt{1998ApJ...501...69D}). The $\chi^{2}$ test examines whether the observed scatter of a light curve is consistent with the reported photometric uncertainties, while the $F$-test evaluates the variance ratio between the target light curve and that of the comparison stars. The ANOVA test probes variability by dividing the light curve into groups and comparing the variance within and between the groups.

For each observing night and each optical band, a conservative criterion was adopted: intraday variability was considered to be detected only when all three statistical tests rejected the null hypothesis of no variability at the 99\% confidence level. This approach minimizes the risk of spurious detections caused by photometric noise or unfavorable observing conditions. Among the 24 nights of intraday observations, we found that three nights (MJD~58420, 58515, and 58821) exhibited statistically significant variability simultaneously in all three optical bands. Notably, on MJD 60275, the $R$ band data exhibited the most significant IDV amplitude with high precision. Although IDV were failed to be detected in the $B$ and $V$ band data because of the low data quality, this night should be considered to have IDV event. Other cases with sporadic detections in individual bands were not regarded as confirmed IDV.

For the bands classified as variable, the variability amplitude was quantified following (\citet{1996A&A...305...42H}):
\begin{equation}
\mathrm{Amp} = 100\% \times \sqrt{(A_{\max} - A_{\min})^{2} - 2\sigma^{2}},
\end{equation}
where $A_{\max}$ and $A_{\min}$ are the maximum and minimum magnitudes of the calibrated light curve, respectively, and $\sigma$ is the mean photometric uncertainty. The smallest amplitude is detected on MJD~58515 in the $V$ band, with a value of 3.88\%, while the largest amplitude amounts to 29.90\% in the $R$ band on MJD~60275. For other nights, the variability amplitudes are $\lesssim 11$\%, which is consistent with previous intraday variability studies of TXS~0506+056(\citealt{2021BlgAJ..34...79B,2024MNRAS.527.1344D}). Overall, our observations generally follow the expected frequency-dependent trend, showing larger variability amplitudes at higher frequencies. Cases where the variability at lower frequencies is comparable to or even larger than that at higher frequencies have also been reported in previous studies (\citealt{2000ApJ...537..638G,2015MNRAS.452.4263G}), possibly due to the dilution of the high-frequency emission by a stable accretion disk or geometric effects within the jet (\citealt{2012ApJ...756...13B, 2015MNRAS.452.4263G}).
The results of the statistical tests and the corresponding variability amplitudes are summarized in Table~\ref{Tab1}.

\subsection{Color Behavior}
\begin{figure*}[t!]
\centering
\includegraphics[width=\textwidth]{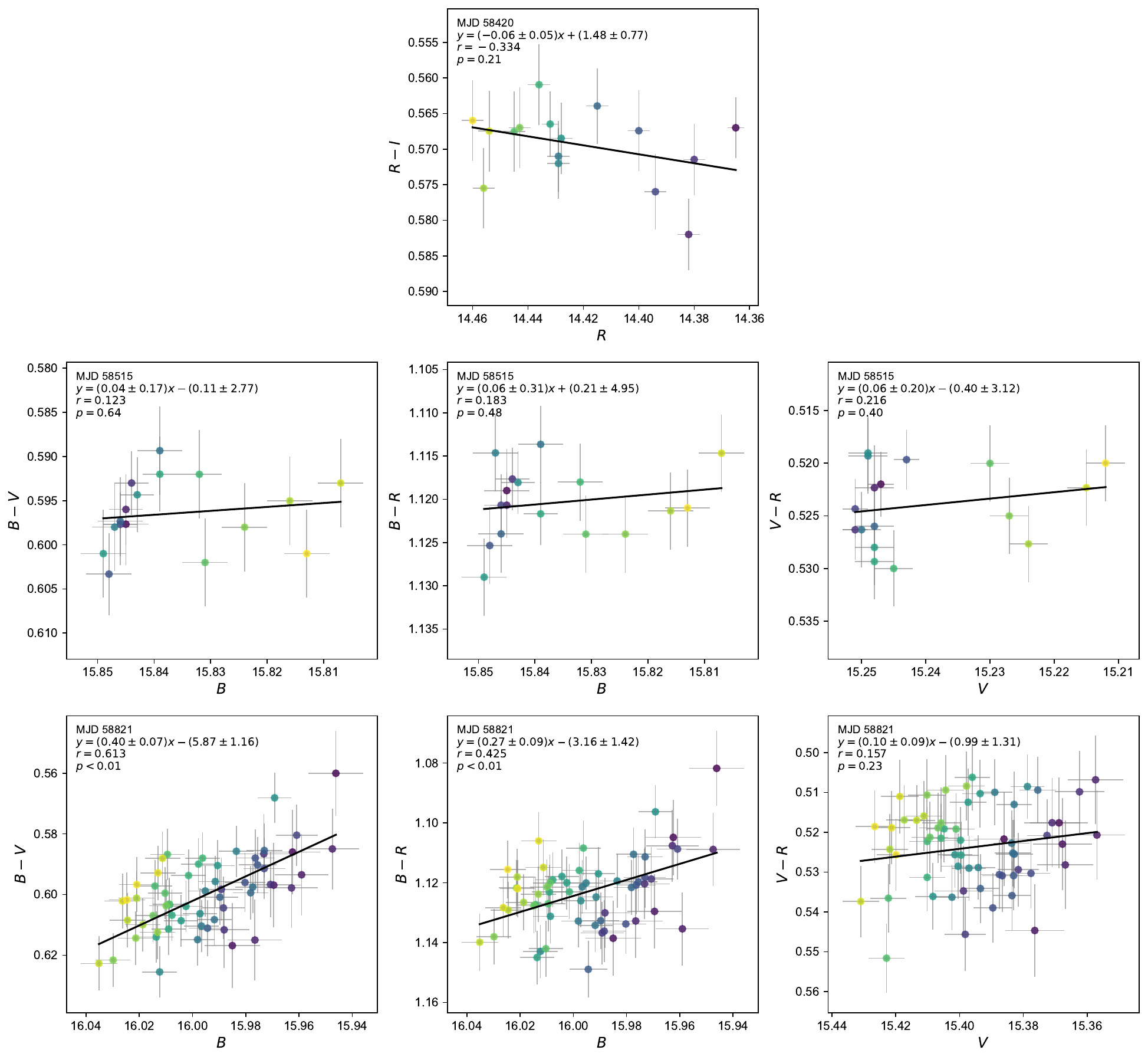}
\caption{CMDs of TXS~0506+056 for the nights exhibiting IDV. The colors from cold to warm indicate the temporal sequence.}
\label{Fig3}
\end{figure*}

For the night on which intraday variability was detected, we investigated the optical color behavior of TXS~0506+056 by means of CMDs. The quasi-simultaneous intraday light curves were linearly interpolated to construct paired magnitudes, from which the $B-V$, $B-R$, $V-R$, and $R-I$ color indices (CIs) were calculated. To quantify the relationship between color and brightness, we performed linear regression between the CIs and the corresponding magnitudes using the bivariate correlated errors and intrinsic scatter (BCES) estimator (\citealt{1996ApJ...470..706A,2012Sci...338.1445N}), which properly accounts for measurement uncertainties in both variables. The significance of the correlation was evaluated using the Spearman rank correlation coefficient. A reliable color--magnitude correlation was considered to be present only when the absolute value of the correlation coefficient ($r$) exceeded 0.2 and the statistical significance was greater than 99\% (i.e., $p < 0.01$). The resulting CMDs for the nights showing variability are shown in Figure~\ref{Fig3}. 

Significant positive correlations are detected in the $B-V$ versus $B$ and $B-R$ versus $B$ diagrams on MJD~58821, indicating a clear BWB trend. In contrast, a weak RWB tendency is noticed on MJD 58420 (with $r < -0.2$ but a relatively large $p$-value). Other variable nights generally exhibit no statistically significant correlations, suggesting approximately achromatic behaviors on intraday timescales, as also reported in, e.g., \citet{2024MNRAS.528.4702M}. 

The observed BWB trend is consistent with previous studies of TXS~0506+056 and other blazars (\citealt{2007AJ....133.1599W,2017MNRAS.465.4423G,2020ApJ...902...41W,2024MNRAS.527.1344D,2026ApJ..1001...32X}). According to the shock-in-jet model developed in \citet{1985ApJ...298..114M,1998A&A...333..452K,2005AJ....130.1418J,2014ApJ...780...87M}, a BWB trend occurs where the electron acceleration timescale is significantly shorter than the cooling timescale. When the electron injection rate increases, high-energy electrons responsible for the high-frequency synchrotron radiation are produced very rapidly, leading to a hardening of the injected electron energy distribution (EED). This results in the high-frequency flux rising faster than the low-frequency flux, causing a BWB trend. On the other hand, the RWB trend on MJD~58420 may be attributed to the intrinsic FSRQ nature of TXS 0506+056. Although emission lines are weak, the less-variable components may contribute a constant flux to certain bands and alter the color-magnitude relationship, as suggested by \citet{2011MNRAS.418.1640W}. The observed achromatic behavior suggests that the short-term variations are driven by changes in the Doppler factor, likely resulting from a varying viewing angle to the emission region \citep[e.g.,][]{2002A&A...390..407V,2012MNRAS.425.3002G,2016MNRAS.455..680A,2023MNRAS.522..102R}.

It is worth noting that no more complex color–spectral behaviors, such as color reversal (i.e., a transition from a BWB to a RWB trend, or vice versa) or spectral hysteresis loops, were detected in our data for this night, although both phenomena have been reported on intraday timescales in BL Lacertae \citep[e.g.,][]{2022ApJ...926...91F,2026ApJ..1001...32X}. The theoretical interpretation links the pattern of spectral hysteresis to the interplay between the electron acceleration and cooling timescales \citep{1998A&A...333..452K}. The competition between these two processes dictates the direction of the loop: a clockwise rotation (a ``soft lag'') is expected when the acceleration process is significantly faster than the cooling process, whereas a counterclockwise rotation (a ``hard lag'') is produced when the two timescales are of a similar order. The absence of these features in our observation of TXS~0506+056 suggests that the relationship between the acceleration and cooling timescales remained relatively stable across the observed bands during the observed variability.

\subsection{Cross-correlation Analysis}

To investigate possible interband time delays, we performed a cross-correlation analysis of the variability observed at different wavelengths using the interpolated cross-correlation function (ICCF; \citealt{1998ApJ...501...82P,1998PASP..110..660P}). A lag search range of $\pm$60 minutes was adopted throughout the analysis.

The ICCF analysis was carried out with the publicly available \texttt{PyCCF} package\footnote{\url{https://ascl.net/code/v/1868}}. The lag uncertainties were estimated from 10,000 Monte Carlo realizations to construct the cross-correlation centroid distribution (CCCD). Since the CCCD may exhibit multiple peaks, we applied the alias removal procedure of \citet{2017ApJ...851...21G} to isolate the primary peak. The final time lag was taken as the median of this peak, with uncertainties defined by the 16th and 84th percentiles. Further details of the Monte Carlo procedure are provided in Appendix~A.

\begin{figure*}[t!]
\centering
\includegraphics[width=\textwidth]{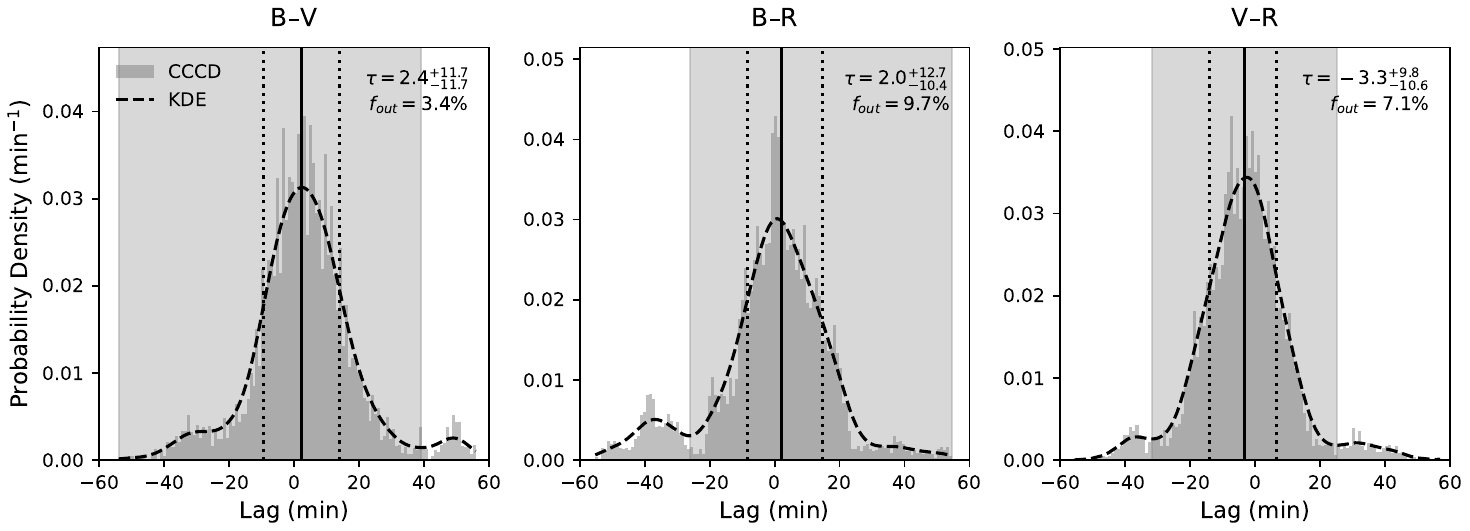} 
\caption{Optical Interband Time Lags for TXS~0506+056 on MJD~58821. The shaded regions indicate the primary peaks of the CCCD after the alias removal procedure.}
\label{Fig4}
\end{figure*}

The results of the ICCF analysis for the variable nights are presented in Table~\ref{Tab2}, with Figure~\ref{Fig4} showing MJD~58821 as a representative example. No significant interband time lag was detected for any of the analyzed nights, as all measured lags are either consistent with zero within their $1\sigma$ uncertainties or too uncertain to support a reliable detection. This null detection is in agreement with our findings in Section~3.2, where no spectral hysteresis loops were observed in the CMDs.

The absence of time lags between variations in different optical bands on intraday timescales is a common result in blazar variability studies \citep[e.g.,][]{2018ApJ...862..123M,2018MNRAS.478.3513Z,2022ApJ...933..224F}. Detecting such lags is intrinsically difficult due to the small wavelength separation between optical bands, which leads to very short expected delays, and the strong dependence on variability amplitude, temporal resolution, and photometric accuracy \citep{2012AJ....143..108W}. The lack of a measurable lag in our data suggests that the emission regions for the $B$, $V$, and $R$ bands are likely co-spatial within the jet, at least to the precision afforded by our observations.

\begin{deluxetable}{cccccr}
\tabletypesize{\footnotesize}
\tablewidth{0pt}

\tablecaption{
Optical Interband Time Lags of IDVs of TXS~0506+056
\label{Tab2}
}

\tablehead{
\colhead{MJD} &
\colhead{Bands} &
\colhead{$f_{\rm fail}$ (\%)} &
\colhead{$f_{\rm out}$ (\%)} &
\colhead{$r_{\rm max}$} &
\colhead{lag (min)}
}

\startdata
58420 & $R-I$ & 99.87 & 7.7 & 0.98 & $-13^{+15}_{-24}$ \\
58515 & $B-V$ & 20.77 & 0.3 & 0.95 & $2^{+11}_{-12}$ \\
      & $B-R$ & 70.41 & 2.4 & 0.96 & $-11^{+9}_{-11}$ \\
      & $V-R$ & 75.49 & 1.7 & 0.96 & $-6^{+11}_{-14}$ \\
58821 & $B-V$ & 11.75 & 3.4 & 0.88 & $2^{+12}_{-12}$ \\
      & $B-R$ & 19.83 & 9.7 & 0.90 & $2^{+13}_{-10}$ \\
      & $V-R$ & 37.09 & 7.1 & 0.89 & $-3^{+10}_{-11}$ \\
\enddata

\tablecomments{
$f_{\rm fail}$ and $f_{\rm out}$ represent the fractions of failed and outlier realizations, respectively. $r_{\rm max}$ is the maximum correlation coefficient. The uncertainties of time lags are given at the $1\sigma$ confidence level.
}

\end{deluxetable}

\subsection{Long-term Variability}

\begin{figure*}[t!]
\centering
\includegraphics[width=\textwidth]{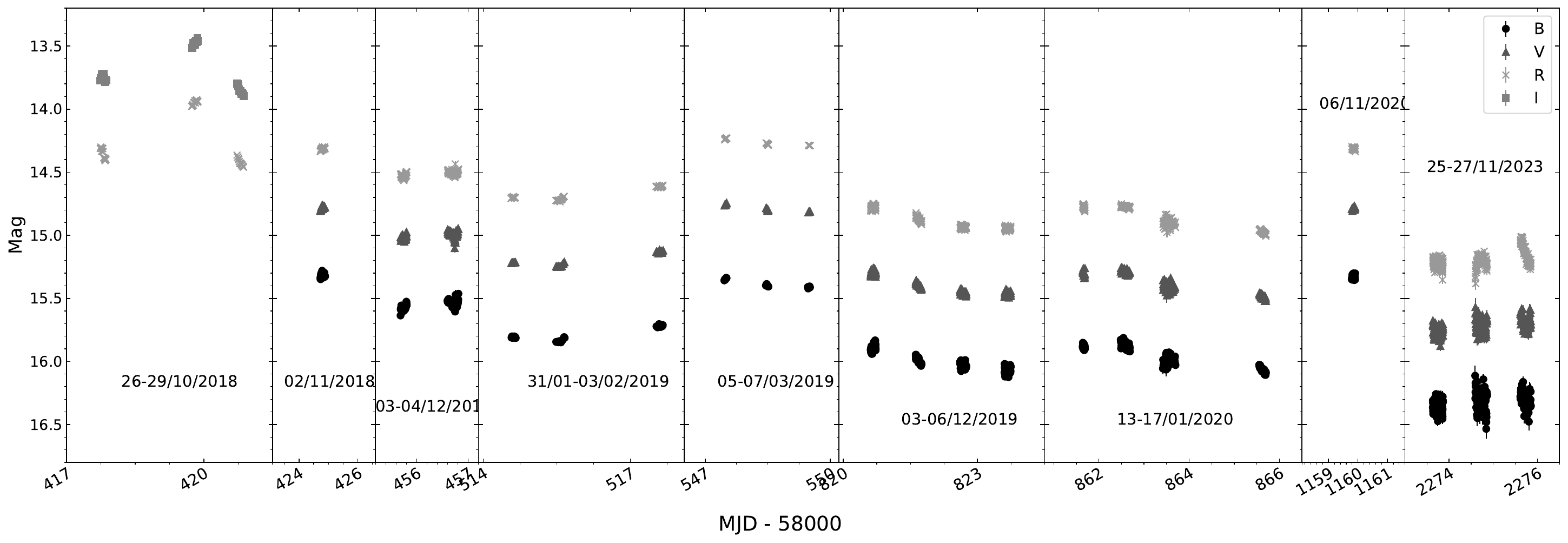} 
\caption{Long-term Optical Light Curves of TXS~0506+056.}
\label{Fig5}
\end{figure*}

The long-term optical light curves observed between October 2018 and November 2023 are presented in Figure~\ref{Fig5}. Throughout the entire monitoring period, the source exhibited an overall dimming trend with several superimposed minor fluctuations, and the flux variations across the $B$, $V$, and $R$ bands are highly correlated. The source reached a peak brightness of $R=13.93$~mag on MJD~58418 and a faintest state of $B=16.27$~mag on MJD~60274, corresponding to a total variability amplitude of approximately 145.6\%.

\begin{figure*}[t!]
\centering
\includegraphics[width=\textwidth]{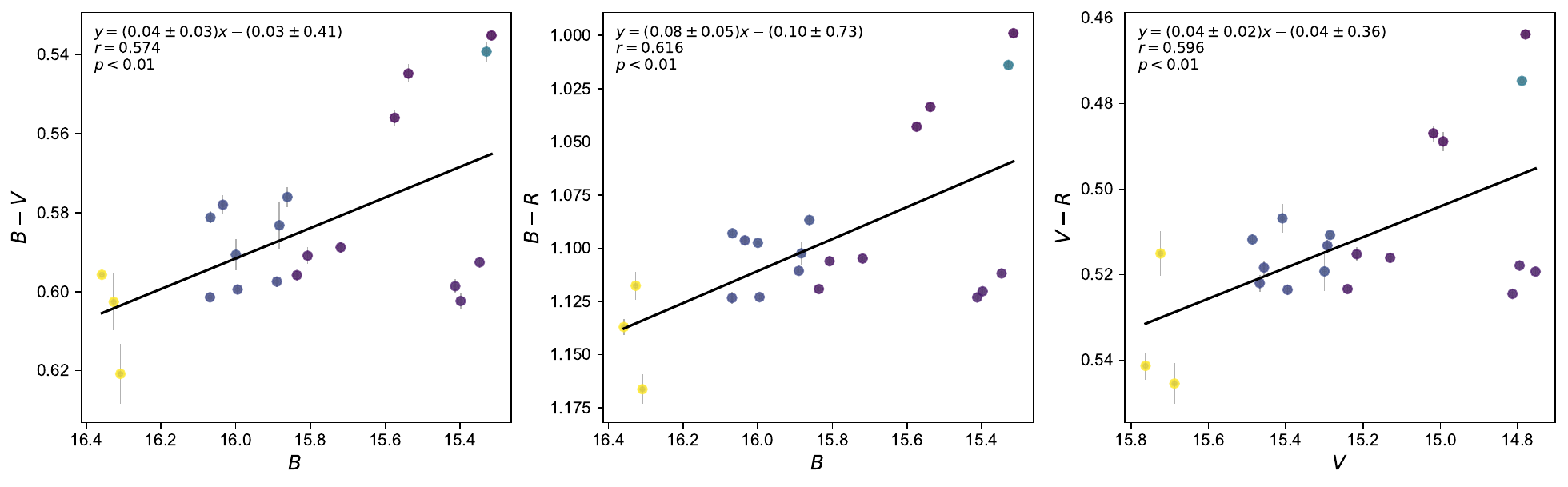} 
\caption{Long-term CMDs of TXS 0506+056. The colors from cold to warm indicate the temporal sequence.}
\label{Fig6}
\end{figure*}

To further investigate the spectral evolution, we constructed CMDs using daily-averaged color indices. As shown in Figure~\ref{Fig6}, the source exhibits a clear BWB trend across all bands, mirroring the findings of \citet{2024MNRAS.527.1344D}. While chromatism in some blazars is more pronounced on short timescales than over longer periods—a behavior often attributed to geometric effects \citep[e.g.,][]{1997A&A...327...61G,2004A&A...421..103V,2021MNRAS.504.5629R}, the correlation observed here suggests that the long-term variability of TXS~0506+056 is likely dominated by intrinsic energetic processes, such as the shift of the synchrotron peak.

\section{Multi-broadband Variability and SED Modeling}
\label{sect:analysis}

\subsection{Long-term Variability}

\begin{figure*}[t!]
\centering
\includegraphics[width=\textwidth]{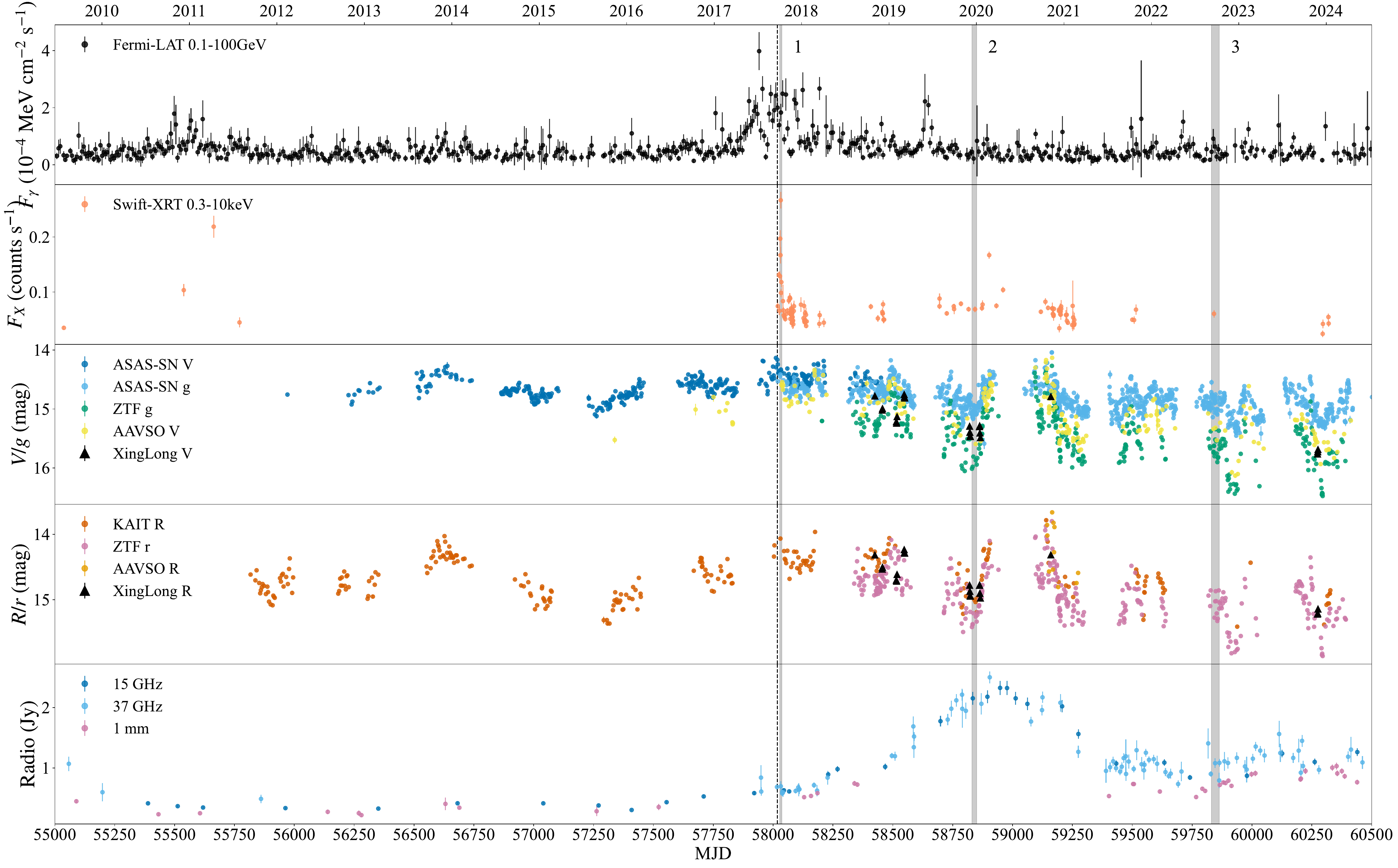} 
\caption{Long-term Multi-broadband Light Curves of TXS~0506+056. The black dashed line marks the observation time of the IC--170922A, and the shaded regions indicate the epochs selected for SED modeling.}
\label{Fig7}
\end{figure*}

Figure~\ref{Fig7} displays the long-term multi-broadband light curves of TXS~0506+056, spanning the period from MJD~55000 to MJD~60500. This dataset incorporates observations from various facilities introduced in Section~\ref{sec:obs}, augmented by the nightly averaged fluxes derived from our 24 intraday observations described in Section~2.1. The multi-broadband light curves reveal a complex and irregular variability pattern over this approximately 15-year period.

An initial flare was evident between MJD~55450 and MJD~55650. During this interval, the $\gamma$-ray flux reached a level of $1.79 \times 10^{-4} \, \mathrm{MeV \, cm^{-2} \, s^{-1}}$ , which was approximately 6 times its quiescent level. The concurrent X-ray data showed a similar trend. However, the scarcity of the optical and radio data prevented revealing a flaring activity. Following this event, the $\gamma$-ray emission largely returned to a prolonged quiescent state, interspersed with some minor flares. The X-ray data were not available and the sparse radio observations limited the characterization of its emission. The optical band, in contrast, exhibited a fluctuating state, notably featuring a flare during MJD~56550--56750 that coincides with a $\gamma$-ray flux enhancement. The period of activity around 2014-2015 is particularly interesting as it coincides with the archival neutrino excess reported by \citet{2018Sci...361..147I}.

Around MJD~57700, the source commenced a phase of intense $\gamma$-ray flaring, coinciding with the high-energy neutrino event IceCube-170922A \citep{2018Sci...361.1378I}. The $\gamma$-ray flux reached a peak of $3.98 \times 10^{-4} \, \mathrm{MeV \, cm^{-2} \, s^{-1}}$ on MJD~57942, representing a more than $10$ times increase relative to the quiescent state. Concurrently, Swift-XRT observations recorded a peak flux of $0.27 \, \mathrm{counts \,s^{-1}}$, consistent with the broader X-ray monitoring reported in \citet{2018ApJ...864...84K}. While the optical flux remained in a sustained high state, the radio emission was at a low flux level but exhibited an increasing trend. The intense $\gamma$-ray flare continued until MJD~58600, while the X-ray emission reverted to a quiescent state relatively soon after the neutrino detection. A secondary $\gamma$-ray flare was detected around MJD~58650, but simultaneous observations in other bands were unavailable for comparison. Subsequently, the $\gamma$-ray emission again settled into a quiescent period, occasionally exhibiting small, random flares.

Notably, after the neutrino event, the radio emission initiated a long-timescale flaring episode, reaching its peak flux density of $\sim$2.3--2.5~Jy around MJD~58900 \citep[e.g.,][]{2021A&A...650A..83H,2026arXiv260401196S}. Both the 15~GHz and 37~GHz observations showed consistent variations during this flare, while the 1~mm observations were unavailable. This radio flare persisted until approximately MJD~59750, after which a secondary, smaller flare commenced.

\subsection{Cross-correlation Analysis}

\begin{figure*}[t!]
\centering
\includegraphics[width=\textwidth]{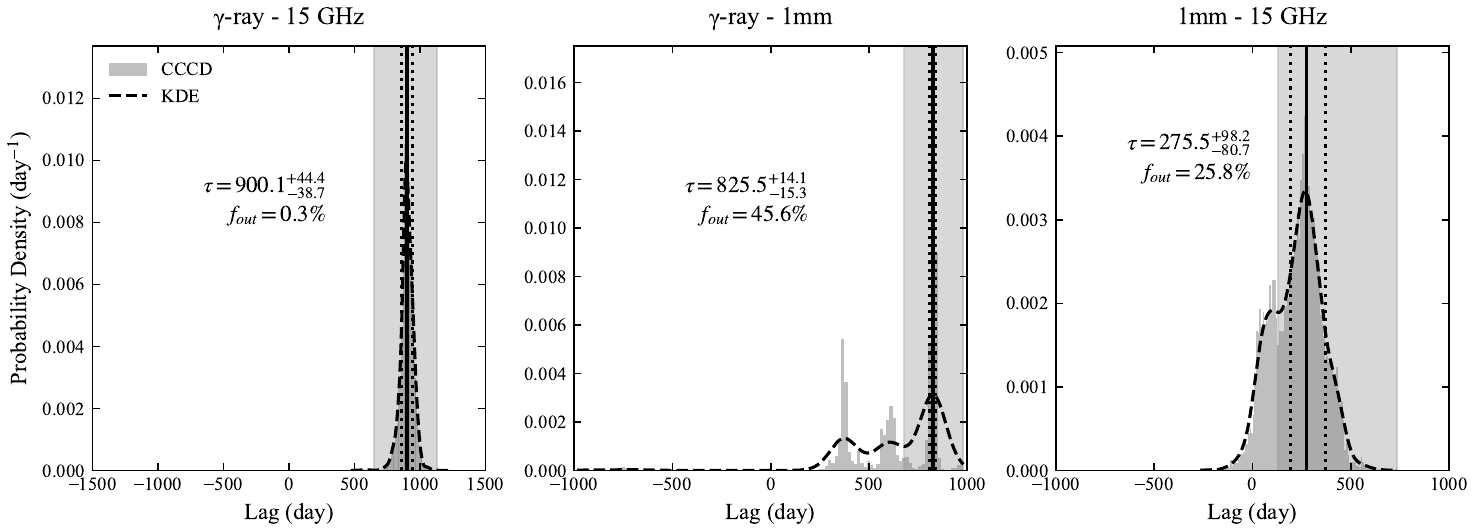}
\caption{Examples of Multi-broadband Time Lags Results. The legends are the same as in Figure~\ref{Fig4}.}
\label{Fig8}
\end{figure*}

As suggested by the light curves in the previous section, the flaring activities across different wavebands may be correlated with potential time delays, which can provide insights into the relative locations of their corresponding emission regions within the jet. To quantify these relationships, we employed the cross-correlation analysis detailed in Section~3.3. Extensive tests were conducted using various data segments and search ranges (e.g., $\pm$500, $\pm$1000, and $\pm$1500 days) to robustly probe for inter-band time lags. Some results are presented in Table~\ref{Tab3}, with representative examples of the correlation analysis shown in Figure~\ref{Fig8}.

\begin{deluxetable}{rclcccc}
\tabletypesize{\footnotesize}
\tablewidth{0pt}

\tablecaption{Multi-broadband Time Lag Results of TXS~0506+056 \label{Tab3}}

\tablehead{
\multicolumn{3}{c}{Bands} &
\colhead{$f_{\rm fail}$ (\%)} &
\colhead{$f_{\rm out}$ (\%)} &
\colhead{$r_{\rm max}$} &
\colhead{lag (day)}
}

\startdata
$\gamma$-ray & -- & X-ray
 & 20.01 & 32.3 & 0.30
 & $-170^{+71}_{-14}$ \\
$\gamma$-ray & -- & V
 & 14.36 & 0.4 & 0.55
 & $9^{+6}_{-6}$ \\
$\gamma$-ray & -- & 1\,mm
 & 20.75 & 45.6 & 0.26
 & $826^{+14}_{-15}$ \\
$\gamma$-ray & -- & 37\,GHz
 & 0.0 & 0.4 & 0.66
 & $834^{+30}_{-44}$ \\
$\gamma$-ray & -- & 15\,GHz
 & 0.0 & 0.3 & 0.69
 & $900^{+44}_{-39}$ \\
1\,mm & -- & 15\,GHz
 & 16.61 & 25.8 & 0.78
 & $276^{+98}_{-81}$ \\
37\,GHz & -- & 15\,GHz
 & 0.0 & 0.0 & 0.96
 & $55^{+43}_{-33}$ \\
\enddata
\end{deluxetable}

Although non-zero lag estimates were obtained for the $\gamma$-ray emission relative to some optical bands and the X-ray band, these lags were not consistently recovered across different data segments or search ranges. Combined with the lack of X-ray observations during part of the monitoring period, the results provide no robust evidence for significant interband time delays and are consistent with emissions from the $\gamma$-ray to optical bands originating from a co-spatial region.

In contrast, a significant and long time lag between the high-energy and radio bands was detected. Our analysis reveals a delay of approximately 800--900 days of the radio emission relative to the $\gamma$-ray outburst. This long delay is robustly determined, as the time lags across the $\gamma$-ray, 1~mm, and 15~GHz bands form a coherent sequence. We note, however, that the absence of 1~mm observations during the period MJD~58200--59300 results in relatively weak constraints on the time lags involving the 1~mm band. Our result differs from the $\sim$ 100-day lag (optical leading radio) reported by \cite{2025Univ...11..204M} and \cite{2018MNRAS.480..192P} based on the Owens Valley Radio Observatory (OVRO) data prior to MJD~58000, which did not cover the prominent outburst after that. So their correlation analysis led to the much shorter lag.

Similar delays have been increasingly recognized in blazar studies \citep[e.g.,][]{2023Symm...15..270K,2026arXiv260403847K} and indicate that the radio emission originates from a distinct, downstream region in the jet. \cite{2026arXiv260401196S} argued that adiabatic expansion alone could not explain the $\sim$2--3 year lag, and they favored instead a re-dissipation event at parsec scales. Furthermore, the cross-correlation analysis among the radio bands (1~mm, 37~GHz, and 15~GHz) reveals a trend where variations at high frequencies lead those at low frequencies, which is a classic signature of a shock wave propagating along the relativistic jet \citep[e.g.,][]{1985ApJ...298..114M}.

\subsection{SED Modeling}
\subsubsection{Epoch Selection and High-energy Spectral Properties}

To investigate the physical conditions and radiation mechanisms within the source, we selected three distinct epochs for detailed broad-band SED modeling. These epochs, highlighted as shaded regions in Figure~\ref{Fig7}, represent contrasting states of the blazar's activity. Epoch 1, spanning MJD~58027--58036, corresponds to the prominent $\gamma$-ray flare associated with the IceCube neutrino event. During this period, the source was also in a flaring state in the X-ray and optical bands, while the radio emission remained at a low level. Epoch 2, from MJD~58830 to MJD~58850, represents a post-flare phase where the $\gamma$-ray, X-ray, and optical emission returned to a quiescent state. Notably, this second period coincides with the peak of a major radio outburst. Epoch 3 (MJD~59830--59864) corresponds to a quiescent state across all wavebands. These epochs were specifically chosen due to their good multi-wavelength data coverage, which is essential for constraining the SED models.

\begin{figure}[ht!]
\centering
\includegraphics[width=\columnwidth]{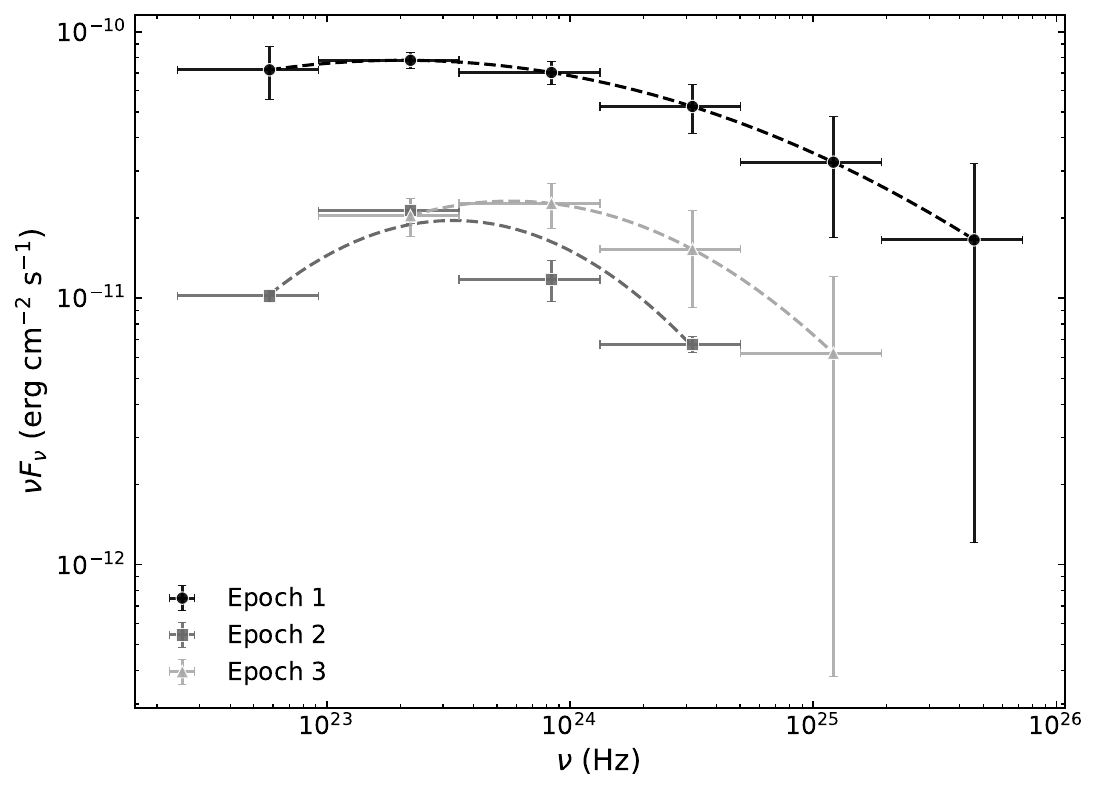}
\caption{Fermi-LAT Observation SED Fitting.}
\label{Fig9}
\end{figure}

\begin{figure*}[t!]
\centering
\includegraphics[width=\textwidth]{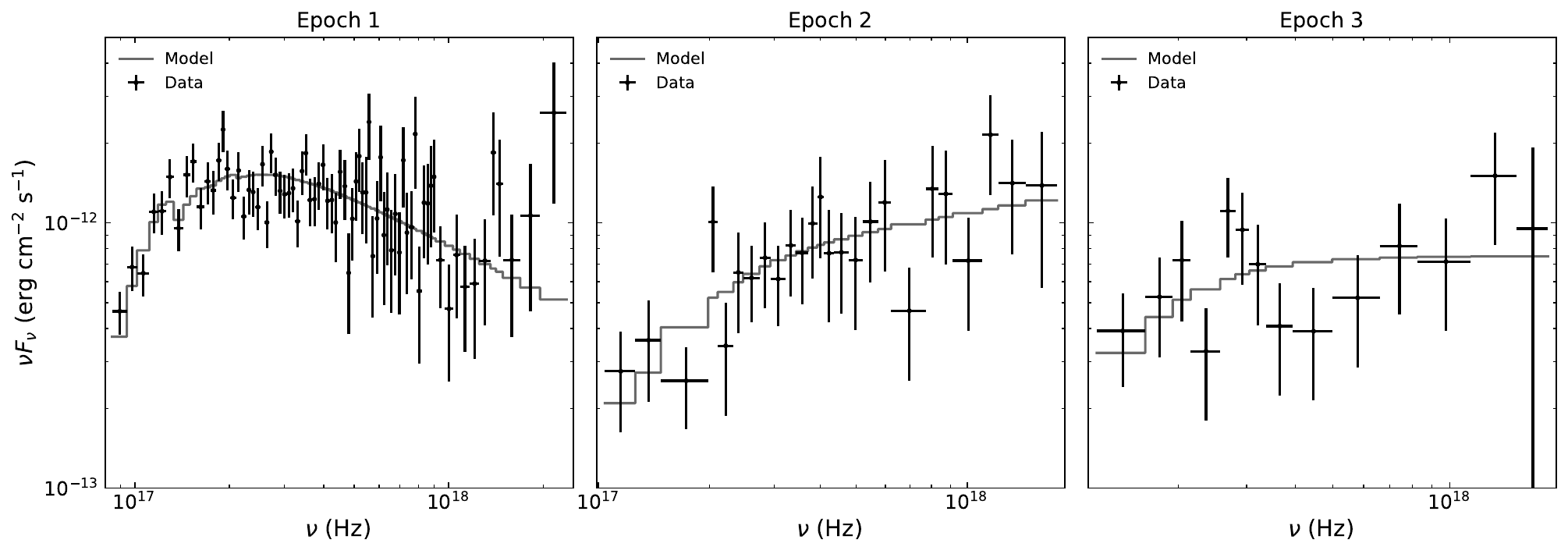} 
\caption{Swift-XRT Observation SED Fitting.}
\label{Fig10}
\end{figure*}

All optical observations were corrected for Galactic extinction. The high-energy spectra were extracted following the procedures described in Sections~2.2.2 and 2.2.3, including the correction for Galactic absorption in the X-ray spectral analysis. The resulting time-averaged $\gamma$-ray and X-ray spectra for both periods are presented in Figures~\ref{Fig9} and ~\ref{Fig10}, respectively, with their analysis results summarized in Tables~\ref{Tab4} and ~\ref{Tab5}.

\begin{deluxetable*}{ccrcr}
\tabletypesize{\footnotesize}
\tablewidth{540pt}
\tablecaption{Analysis Results of Fermi-LAT Observations \label{Tab4}}
\tablehead{
\colhead{Epoch} & 
\colhead{$\alpha$} & 
\colhead{$\beta$} & 
\colhead{$F_{\gamma}$ ($\rm erg\,cm^{-2}\,s^{-1}$)} & 
\colhead{TS}
}
\startdata
1 & $2.05\pm0.08$ & $0.054\pm0.05$ & $(3.22\pm0.52)\times10^{-10}$ & 743.3 \\
2 & $2.22\pm0.01$ & $0.423\pm0.01$ & $(4.50\pm0.14)\times10^{-11}$ & 71.5  \\
3 & $1.82\pm0.15$ & $0.147\pm0.09$ & $(7.72\pm1.06)\times10^{-11}$ & 167.4  \\
\enddata

\tablecomments{The columns list the two parameters of the log-parabola spectral model, the integrated $\gamma$-ray flux in the energy range of 0.1--300~GeV, and the maximum-likelihood test statistic.
}
\end{deluxetable*}

\begin{deluxetable}{ccrcc}
\tabletypesize{\scriptsize}   
\tablewidth{0pt}
\tablecaption{Analysis Results of Swift-XRT Observations \label{Tab5}}
\tablehead{
\colhead{Epoch} & 
\colhead{Exp. (ks)} & 
\colhead{$\Gamma_{\rm X}$} &
\colhead{$F_{\rm X}$ ($10^{-12}\,\rm erg\,cm^{-2}\,s^{-1}$)} & 
\colhead{$\chi^{2}_{\rm red}$}
}
\startdata
1 & 14.9 & $2.62^{+0.08}_{-0.07}$ & $5.99^{+0.28}_{-0.27}$ & 0.91 \\
2 & 3.2  & $1.77^{+0.23}_{-0.20}$ & $3.20^{+0.50}_{-0.40}$ & 0.83 \\
3 & 2.5  & $2.00^{+0.39}_{-0.27}$ & $2.70^{+0.50}_{-0.50}$ & 0.92 \\
\enddata

\tablecomments{The columns list the X-ray exposure time, the best-fit photon index of the power-law model, the integrated X-ray flux in the 0.3--10~keV energy range, and the reduced $\chi^{2}$.
}
\end{deluxetable}

Our spectral analysis results are broadly consistent with previous studies utilizing different datasets and methodologies \citep[e.g.,][]{2019ApJ...880..103G,2022A&A...668A.146D,2024ApJ...962..142W}. A comparison across the three epochs reveals clear spectral evolution. In the $\gamma$-ray band, the flux dropped from $3.22\times10^{-10}$ in Epoch 1 to $4.50\times10^{-11}\,\rm erg\,cm^{-2}\,s^{-1}$ in Epoch 2, with the spectral index $\alpha$ exhibiting variations between 2.05 and 2.22. Conversely, the X-ray band shows a "softer-when-brighter" behavior: the photon index $\Gamma_{\rm X}$ increased from $1.77$ in Epoch 2 to $2.62$ in Epoch 1 as the flux nearly doubled. The softening in the X-ray spectrum during the flare likely reflects the injection of fresh, lower-energy electrons.

\subsubsection{SED Modeling and Results}

We modeled the broad-band SEDs of the three selected epochs using a time-dependent, one-zone lepto-hadronic model. The numerical simulations are performed with the AM$^3$ code\footnote{\url{https://am3.readthedocs.io/en/latest/index.html}} \citep{2024ApJS..275....4K}, which self-consistently treats key radiation processes, including synchrotron, inverse compton, and various hadronic interactions (e.g., photo-pion production, Bethe-Heitler pair production, $\gamma\gamma$ annihilation). The model assumes a spherical emission region (a `blob') of radius $R'$, moving with a bulk Lorentz factor $\Gamma$ at a small angle to our line of sight. Relativistic electrons are injected with a broken power-law energy distribution, while protons follow a single power-law.

\begin{figure*}[t!]
\centering
\includegraphics[width=0.95\textwidth]{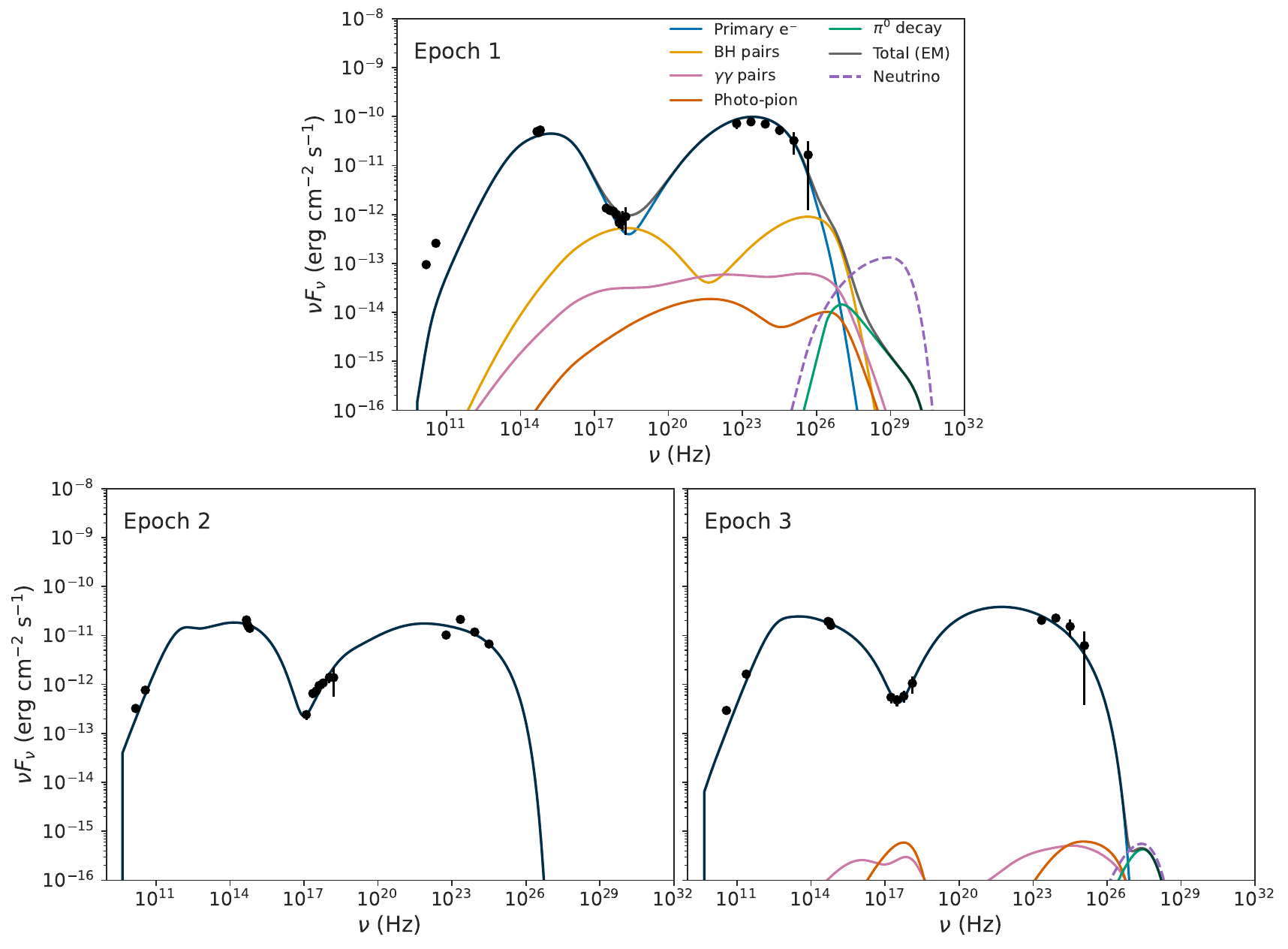}
\caption{SED modelings of TXS~0506+056. The black points denote the contemporaneous multiwavelength data. The model components are synchrotron and inverse-Compton emission from the primary electrons (blue), BH pairs (orange), secondary pairs from $\gamma\gamma$ interactions (pink) and photohadronic processes (red), and $\pi^{0}$ decay (green). The gray curve shows the integrated emission, and the purple dashed curve, the neutrino spectrum. For Epoch~2, some components are omitted due to their negligible radiative contributions.}
\label{Fig11}
\end{figure*}

The free parameters of the model, including $\Gamma$, magnetic field strength $B'$, blob radius $R'$, injected luminosities ($L'_e, L'_p$), and particle energy parameters, were constrained by fitting the model output to the quasi-simultaneous observational data from optical to $\gamma$-rays. To guide the fitting process and ensure a physically meaningful search, the initial values and allowed ranges for these parameters were set based on the typical values reported in previous modelings of this source \citep[e.g.,][]{2019NatAs...3...88G,2019ApJ...874L..29R}. The best-fit parameters were obtained by minimizing the $\chi^2$ statistic. Following the standard blazar jet paradigm, the radio data were excluded from the fit, as the low-frequency emission is generally attributed to a larger, more downstream jet region distinct from the compact high-energy emission site \citep[e.g.,][]{2020ApJ...891..115P,2022ApJ...927..197A}. The resulting best-fit SEDs for three epochs are shown in Figure~\ref{Fig11}, and the corresponding physical parameters are summarized in Table~\ref{Tab6}.

\begin{deluxetable}{lccc}
\tabletypesize{\footnotesize}
\tablewidth{0pt}

\tablecaption{Best-fit parameters for SED model of TXS~0506+056 \label{Tab6}}

\tablehead{
\colhead{Parameter} & 
\colhead{Epoch 1} & 
\colhead{Epoch 2} & 
\colhead{Epoch 3}
}

\startdata
\tableline
$z$ & 0.337 & 0.337 & 0.337 \\
$\gamma'_{p,\rm min}$ & 10 & 10 & 10 \\
$\alpha_p$ & $-2.0$ & $-2.0$ & $-2.0$ \\
$\eta_{\rm esc}$ & $c/10$ & $c/10$ & $c/10$ \\ 
\tableline
$B'$ (G) & 0.13 & 0.0077 & 0.032 \\
$R'$ (cm) & $10^{16.1}$ & $10^{17.8}$ & $10^{17.0}$ \\
$\Gamma$ & 27.8 & 18.7 & 20.0 \\
$L'_{e,\rm inj}$ (erg s$^{-1}$) & $10^{40.6}$ & $10^{41.4}$ & $10^{40.9}$ \\
$\alpha_{e,1}$ & $-1.9$ & $-3.4$ & $-2.7$ \\
$\alpha_{e,2}$ & $-4.2$ & $-2.5$ & $-3.2$ \\
$\gamma'_{e,\rm min}$ & $10^{3.4}$ & $10^{3.1}$ & $10^{3.1}$ \\
$\gamma'_{e,\rm brk}$ & $10^{4.5}$ & $10^{3.5}$ & $10^{3.6}$ \\
$\gamma'_{e,\rm max}$ & $10^{5.0}$ & $10^{4.6}$ & $10^{4.8}$ \\
$L'_{p,\rm inj}$ (erg s$^{-1}$) & $10^{45.8}$ & $10^{43.0}$ & $10^{44.0}$ \\
$\gamma'_{p,\rm max}$ & $10^{5.8}$ & $10^{3.5}$ & $10^{4.2}$ \\
\enddata

\tablecomments{The first four are fixed parameters (above the horizontal line), while the others are free parameters in the SED fitting. The primed quantities refer to the rest frame of the radiation zone (blob). $\eta_{\rm esc}$ parameterizes the escape rate of $e^\pm$ and $p$. $B'$, $R'$, and $\Gamma$ denote the magnetic field strength, emitting region radius, and bulk Lorentz factor, respectively. $L'_{e,\rm inj}$ and $L'_{p,\rm inj}$ are the injected electron and proton luminosities, while $\gamma'$ and $\alpha$ represent the particle Lorentz factors and spectral indices.
}
\end{deluxetable}

For Epoch~1, the modeling indicates that the X-ray emission is predominantly driven by hadronic processes. This scenario, characterized by a high proton luminosity ($L'_{p,\rm inj} = 10^{45.8}$~erg~s$^{-1}$), naturally accounts for the significant predicted neutrino flux, in line with the findings of other lepto-hadronic models for this source \citep{2018ApJ...864...84K, 2019MNRAS.483L..12C, 2019NatAs...3...88G}. As expected, the single-zone model for the compact flare region underestimates the radio flux, with the observed radio data points lying significantly above the model curve. This radio excess can be attributed to the contribution from extended jet regions downstream.

In contrast, the SEDs of Epoch~2 and 3 are well-reproduced by lepton-dominated lepto-hadronic scenarios, in which the radiative output is primarily contributed by leptonic processes while the hadronic components remain weak. This behavior is naturally explained by the lower magnetic field strength and larger emission region inferred from the SED fitting, which  reduce the radiative efficiency of hadronic processes, allowing the observed SEDs to be reproduced primarily by leptonic emission. Consequently, the predicted neutrino flux for this period is lower than that of Epoch~1, consistent with the observation. Interestingly, the models for these two epochs provide reasonable approximations of the radio data. This is consistent with the fitted physical parameters: the emission region radius $R'$ increases from $10^{16.1}$~cm in Epoch~1 to $10^{17.8}$~cm in Epoch~2, accompanied by a sharp decrease in the magnetic field strength $B'$ from 0.13~G to 0.0077~G. These properties suggest that the emission in Epochs~2 and 3 originates from larger and more downstream regions of the jet, where the plasma becomes optically thin to radio emission. Overall, our multi-epoch modeling results are consistent with the general multi-wavelength and multi-messenger characteristics of the source reported in the literature \citep[e.g.,][]{2019ApJ...874L..29R, 2020ApJ...891..115P, 2022ApJ...927..197A}.

\section{Conclusions} \label{sect:conclusion}

In this work, we carried out optical monitoring and performed a comprehensive, intraday and long-term multiwavelength study of the neutrino-blazar candidate TXS~0506+056, utilizing data spanning from radio to $\gamma$-ray bands. We conducted a detailed analysis of its optical variability on both intraday and long-term timescales, investigated the inter-band correlations and time delays, and performed time-dependent lepto-hadronic modeling of SEDs during three distinct activity states. Our main findings are summarized as follows:

\begin{itemize}
    \item The optical light curves by our observations exhibited an overall dimming trend with superimposed minor fluctuations throughout the monitoring period. IDV was detected on four nights (MJDs~58420, 58515, 58821, and 60275). A clear BWB trend was observed on one night, characteristic of synchrotron emission processes, while another night showed a weak RWB behavior. The cross-correlation analysis revealed no significant inter-band time lags, suggesting a co-spatial emission region for the different optical bands.

    \item The long-term multi-broadband light curves show complex and irregular variability, characterized by asynchronous flaring activities primarily between the radio and high-energy bands. Cross-correlation analysis reveals that emissions from $\gamma$-ray to optical bands are contemporaneous, consistent with a co-spatial origin within the jet. In contrast, a significant delay of $\sim$800--900 days is detected, with the radio emission lagging behind the high-energy emission. This result strongly supports a scenario with at least two distinct emission zones: a compact, upstream region responsible for the high-energy flares, and a larger, downstream region for the radio outbursts. This is further corroborated by the frequency stratification observed within the radio bands.

    \item The SED of the 2017 flaring state (Epoch 1), which was contemporaneous with the IceCube-170922A neutrino event, is well-reproduced by a lepto-hadronic model. The X-ray emission is hadronically dominated, and the high proton luminosity provides a self-consistent physical basis for the associated high-energy neutrino detection.

    \item The post-flare and quiescent states (Epochs 2 and 3) are, in contrast, well-reproduced by leptonically-dominated scenarios, leading to lower neutrino fluxes. The model for these epochs indicates substantially larger emission regions with much weaker magnetic fields compared to those of the flaring state. This suggests that the post-flare and quiescent emissions arise from more extended and downstream regions of the jet, allowing for a larger contribution to the observed radio emission.
    
\end{itemize}

Our detailed multi-epoch and multi-wavelength analyses of TXS~0506+056 described a dynamic picture on this object and its jet physics. Our results suggest a transition from a compact, magnetically dominated, and proton-loaded emission region during the multi-messenger flare to more extended and leptonically dominated emission regions during the post-flare and quiescent states. These discoveries underscore the complex interplay between leptonic and hadronic processes and highlight the critical role of long-term, multi-messenger monitoring in deciphering the mechanisms of particle acceleration and emission in blazar jets.

\begin{acknowledgments}
The authors thank the anonymous referee for the constructive comments and suggestions that helped improve this manuscript. This work is supported by the National Key R\&D Programs of China (2021YFA0718500 and 2025YFA1614101) and by the Chinese National Natural Science Foundation grant No. 12333001. We make use of the public Fermi-LAT data and data provided by the UK Swift Science Data Centre at the University of Leicester. We acknowledge observations from AAVSO, ZTF, ASAS-SN, and KAIT. We thank Mark Gurwell for providing SMA data. The Submillimeter Array is operated by the Smithsonian Astrophysical Observatory and the Academia Sinica Institute of Astronomy and Astrophysics. This publication also makes use of data from the Mets\"ahovi Radio Observatory, operated by Aalto University, Finland. We further acknowledge VLBA data from the VLBA-BU Blazar Monitoring Program and the MOJAVE database.
\end{acknowledgments}

\facilities{Fermi, Swift(XRT), AAVSO, ASAS-SN, ZTF, KAIT, SMA, VLBA, Mets\"ahovi}

\software{Photutils \citep{2024zndo..10967176B}, Fermitools \citep{2019ascl.soft05011F}, HEASoft \citep{2014ascl.soft08004N}, BCES \citep{1996ApJ...470..706A, 2012Sci...338.1445N}, PyCCF \citep{1998PASP..110..660P, 2018ascl.soft05032S}, \texttt{JAVELIN} \citep{2011ApJ...735...80Z}, ZDCF \citep{1997ASSL..218..163A, 2013arXiv1302.1508A}, pyPETaL \citep{2024ApJS..272...26S, 2024ascl.soft01004S}, AM$^3$ \citep{2024ApJS..275....4K}}

\appendix

\section{Monte Carlo estimation of lag uncertainties}

The uncertainties of the ICCF lags were estimated using the Monte Carlo procedure implemented in the \texttt{PyCCF} package. We adopted the combined flux randomization and random subset selection (FR/RSS) scheme, in which the fluxes were randomly perturbed according to their photometric uncertainties and the light curves were resampled with replacement. A total of 10,000 realizations were performed for each pair of light curves. The centroid lag was calculated for each realization to construct the cross-correlation centroid distribution (CCCD), and the lag uncertainties were determined from the 16th and 84th percentiles of the resulting distribution.

\bibliography{sample701}{}

@software{2014ascl.soft08004N,
       author = {{Nasa High Energy Astrophysics Science Archive Research Center (Heasarc)}},
        title = "{HEAsoft: Unified Release of FTOOLS and XANADU}",
 howpublished = {Astrophysics Source Code Library, record ascl:1408.004},
         year = 2014,
        month = aug,
          eid = {ascl:1408.004},
archivePrefix = {ascl},
       eprint = {1408.004},
       adsurl = {https://ui.adsabs.harvard.edu/abs/2014ascl.soft08004N}
}

@software{2024ascl.soft01004S,
       author = {{Stone}, Zachary},
        title = "{pyPETaL: A Pipeline for Estimating AGN Time Lags}",
 howpublished = {Astrophysics Source Code Library, record ascl:2401.004},
         year = 2024,
        month = jan,
          eid = {ascl:2401.004},
archivePrefix = {ascl},
       eprint = {2401.004},
       adsurl = {https://ui.adsabs.harvard.edu/abs/2024ascl.soft01004S}
}

@ARTICLE{2024ApJS..272...26S,
       author = {{Shen}, Yue and {Grier}, Catherine J. and {Horne}, Keith and {Stone}, Zachary and {Li}, Jennifer I. and {Yang}, Qian and {Homayouni}, Yasaman and {Trump}, Jonathan R. and {Anderson}, Scott F. and {Brandt}, W.~N. and et al.},
        title = "{The Sloan Digital Sky Survey Reverberation Mapping Project: Key Results}",
      journal = {\apjs},
         year = 2024,
        month = jun,
       volume = {272},
       number = {2},
          eid = {26},
        pages = {26},
          doi = {10.3847/1538-4365/ad3936},
archivePrefix = {arXiv},
       eprint = {2305.01014},
 primaryClass = {astro-ph.GA},
       adsurl = {https://ui.adsabs.harvard.edu/abs/2024ApJS..272...26S}
}

@software{2018ascl.soft05032S,
       author = {{Sun}, Mouyuan and {Grier}, C.~J. and {Peterson}, B.~M.},
        title = "{PyCCF: Python Cross Correlation Function for reverberation mapping studies}",
 howpublished = {Astrophysics Source Code Library, record ascl:1805.032},
         year = 2018,
        month = may,
          eid = {ascl:1805.032},
archivePrefix = {ascl},
       eprint = {1805.032},
       adsurl = {https://ui.adsabs.harvard.edu/abs/2018ascl.soft05032S}
}

@ARTICLE{2021MNRAS.504.5629R,
       author = {{Raiteri}, C.~M. and {Villata}, M. and {Larionov}, V.~M. and {Jorstad}, S.~G. and {Marscher}, A.~P. and {Weaver}, Z.~R. and {Acosta-Pulido}, J.~A. and {Agudo}, I. and {Andreeva}, T. and {Arkharov}, A. and {Bachev}, R. and {Ben{\'\i}tez}, E. and {Berton}, M. and {Bj{\"o}rklund}, I. and {Borman}, G.~A. and {Bozhilov}, V. and {Carnerero}, M.~I. and {Carosati}, D. and {Casadio}, C. and {Chen}, W.~P. and {Damljanovic}, G. and {D'Ammando}, F. and {Escudero}, J. and {Fuentes}, A. and {Giroletti}, M. and {Grishina}, T.~S. and {Gupta}, A.~C. and {Hagen-Thorn}, V.~A. and {Hart}, M. and {Hiriart}, D. and {Hou}, W.-J. and {Ivanov}, D. and {Kim}, J.-Y. and {Kimeridze}, G.~N. and {Konstantopoulou}, C. and {Kopatskaya}, E.~N. and {Kurtanidze}, O.~M. and {Kurtanidze}, S.~O. and {L{\"a}hteenm{\"a}ki}, A. and {Larionova}, E.~G. and {Larionova}, L.~V. and {Marchili}, N. and {Markovic}, G. and {Minev}, M. and {Morozova}, D.~A. and {Myserlis}, I. and {Nakamura}, M. and {Nikiforova}, A.~A. and {Nikolashvili}, M.~G. and {Otero-Santos}, J. and {Ovcharov}, E. and {Pursimo}, T. and {Rahimov}, I. and {Righini}, S. and {Sakamoto}, T. and {Savchenko}, S.~S. and {Semkov}, E.~H. and {Shakhovskoy}, D. and {Sigua}, L.~A. and {Stojanovic}, M. and {Strigachev}, A. and {Thum}, C. and {Tornikoski}, M. and {Traianou}, E. and {Troitskaya}, Y.~V. and {Troitskiy}, I.~S. and {Tsai}, A. and {Valcheva}, A. and {Vasilyev}, A.~A. and {Vince}, O. and {Zaharieva}, E.},
        title = "{The complex variability of blazars: time-scales and periodicity analysis in S4 0954+65}",
      journal = {\mnras},
         year = 2021,
        month = jul,
       volume = {504},
       number = {4},
        pages = {5629-5646},
          doi = {10.1093/mnras/stab1268},
archivePrefix = {arXiv},
       eprint = {2104.15005},
 primaryClass = {astro-ph.HE},
       adsurl = {https://ui.adsabs.harvard.edu/abs/2021MNRAS.504.5629R}
}

@ARTICLE{2004A&A...421..103V,
       author = {{Villata}, M. and {Raiteri}, C.~M. and {Kurtanidze}, O.~M. and {Nikolashvili}, M.~G. and {Ibrahimov}, M.~A. and {Papadakis}, I.~E. and {Tosti}, G. and {Hroch}, F. and {Takalo}, L.~O. and {Sillanp{\"a}{\"a}}, A. and et al.},
        title = "{The WEBT <ASTROBJ>BL Lacertae</ASTROBJ> Campaign 2001 and its extension.  Optical light curves and colour analysis 1994-2002}",
      journal = {\aap},
         year = 2004,
        month = jul,
       volume = {421},
        pages = {103-114},
          doi = {10.1051/0004-6361:20035895},
archivePrefix = {arXiv},
       eprint = {astro-ph/0404155},
 primaryClass = {astro-ph},
       adsurl = {https://ui.adsabs.harvard.edu/abs/2004A&A...421..103V}
}

@ARTICLE{1997A&A...327...61G,
       author = {{Ghisellini}, G. and {Villata}, M. and {Raiteri}, C.~M. and {Bosio}, S. and {de Francesco}, G. and {Latini}, G. and {Maesano}, M. and {Massaro}, E. and {Montagni}, F. and {Nesci}, R. and {Tosti}, G. and {Fiorucci}, M. and {Pian}, E. and {Maraschi}, L. and {Treves}, A. and {Comastri}, A. and {Mignoli}, M.},
        title = "{Optical-IUE observations of the gamma-ray loud BL Lacertae object S5 0716+714: data and interpretation.}",
      journal = {\aap},
         year = 1997,
        month = nov,
       volume = {327},
        pages = {61-71},
          doi = {10.48550/arXiv.astro-ph/9706254},
archivePrefix = {arXiv},
       eprint = {astro-ph/9706254},
 primaryClass = {astro-ph},
       adsurl = {https://ui.adsabs.harvard.edu/abs/1997A&A...327...61G}
}

@ARTICLE{2024MNRAS.528.4702M,
       author = {{McCall}, Callum and {Jermak}, Helen E. and {Steele}, Iain A. and {Kobayashi}, Shiho and {Knapen}, Johan H. and {S{\'a}nchez-Alarc{\'o}n}, Pablo M.},
        title = "{Detection of an intranight optical hard lag with colour variability in blazar PKS 0735+178}",
      journal = {\mnras},
         year = 2024,
        month = mar,
       volume = {528},
       number = {3},
        pages = {4702-4719},
          doi = {10.1093/mnras/stae310},
archivePrefix = {arXiv},
       eprint = {2402.00633},
 primaryClass = {astro-ph.HE},
       adsurl = {https://ui.adsabs.harvard.edu/abs/2024MNRAS.528.4702M}
}

@ARTICLE{2026arXiv260403847K,
       author = {{Kochocki}, Alina and {Kun}, Emma and {Hori}, Sam},
        title = "{Characterizing Gamma-Radio Delayed Flaring Activity from Blazars}",
      journal = {arXiv e-prints},
         year = 2026,
        month = apr,
          eid = {arXiv:2604.03847},
        pages = {arXiv:2604.03847},
archivePrefix = {arXiv},
       eprint = {2604.03847},
 primaryClass = {astro-ph.HE},
       adsurl = {https://ui.adsabs.harvard.edu/abs/2026arXiv260403847K}
}

@ARTICLE{2023Symm...15..270K,
       author = {{Kun}, Emma and {Medveczky}, Attila},
        title = "{Multiwavelength Analysis of the IceCube Neutrino Source Candidate Blazar PKS 1424+240}",
      journal = {Symmetry},
         year = 2023,
        month = jan,
       volume = {15},
       number = {2},
          eid = {270},
        pages = {270},
          doi = {10.3390/sym15020270},
       adsurl = {https://ui.adsabs.harvard.edu/abs/2023Symm...15..270K}
}

@ARTICLE{2026arXiv260401196S,
       author = {{Stathopoulos}, S.~I. and {Yuan}, C. and {Vasilopoulos}, G. and {Testagrossa}, F. and {Karavola}, D. and {Petropoulou}, M. and {Winter}, W.},
        title = "{Delayed Radio Flares in Neutrino-associated Blazars: The Case of TXS 0506+056}",
      journal = {arXiv e-prints},
         year = 2026,
        month = apr,
          eid = {arXiv:2604.01196},
        pages = {arXiv:2604.01196},
archivePrefix = {arXiv},
       eprint = {2604.01196},
 primaryClass = {astro-ph.HE},
       adsurl = {https://ui.adsabs.harvard.edu/abs/2026arXiv260401196S}
}

@ARTICLE{2022A&A...668A.146D,
       author = {{Das}, Saikat and {Gupta}, Nayantara and {Razzaque}, Soebur},
        title = "{Implications of multiwavelength spectrum on cosmic-ray acceleration in blazar TXS 0506+056}",
      journal = {\aap},
         year = 2022,
        month = dec,
       volume = {668},
          eid = {A146},
        pages = {A146},
          doi = {10.1051/0004-6361/202244653},
archivePrefix = {arXiv},
       eprint = {2208.00838},
 primaryClass = {astro-ph.HE},
       adsurl = {https://ui.adsabs.harvard.edu/abs/2022A&A...668A.146D}
}

@ARTICLE{2019ApJ...880..103G,
       author = {{Garrappa}, S. and {Buson}, S. and {Franckowiak}, A. and {Fermi-LAT Collaboration} and {Shappee}, B.~J. and {Beacom}, J.~F. and {Dong}, S. and {Holoien}, T.~W.-S. and {Kochanek}, C.~S. and {Prieto}, J.~L. and {Stanek}, K.~Z. and {Thompson}, T.~A. and {ASAS-SN Collaboration} and {Aartsen}, M.~G. and {Ackermann}, M. and {Adams}, J. and {Aguilar}, J.~A. and {Ahlers}, M. and {Ahrens}, M. and {Alispach}, C. and {Andeen}, K. and {Anderson}, T. and {Ansseau}, I. and {Anton}, G. and {Arg{\"u}elles}, C. and {Auffenberg}, J. and {Axani}, S. and {Backes}, P. and {Bagherpour}, H. and {Bai}, X. and {Barbano}, A. and {Barwick}, S.~W. and {Baum}, V. and {Bay}, R. and {Beatty}, J.~J. and {Becker}, K.-H. and {Becker Tjus}, J. and {BenZvi}, S. and {Berley}, D. and {Bernardini}, E. and {Besson}, D.~Z. and {Binder}, G. and {Bindig}, D. and {Blaufuss}, E. and {Blot}, S. and {Bohm}, C. and {B{\"o}rner}, M. and {B{\"o}ser}, S. and {Botner}, O. and {Bourbeau}, E. and {Bourbeau}, J. and {Bradascio}, F. and {Braun}, J. and {Bretz}, H.-P. and {Bron}, S. and {Brostean-Kaiser}, J. and {Burgman}, A. and {Busse}, R.~S. and {Carver}, T. and {Chen}, C. and {Cheung}, E. and {Chirkin}, D. and {Clark}, K. and {Classen}, L. and {Collin}, G.~H. and {Conrad}, J.~M. and {Coppin}, P. and {Correa}, P. and {Cowen}, D.~F. and {Cross}, R. and {Dave}, P. and {de Andr{\'e}}, J.~P.~A.~M. and {De Clercq}, C. and {DeLaunay}, J.~J. and {Dembinski}, H. and {Deoskar}, K. and {De Ridder}, S. and {Desiati}, P. and {de Vries}, K.~D. and {de Wasseige}, G. and {de With}, M. and {DeYoung}, T. and {Diaz}, A. and {D{\'\i}az-V{\'e}lez}, J.~C. and {Dujmovic}, H. and {Dunkman}, M. and {Dvorak}, E. and {Eberhardt}, B. and {Ehrhardt}, T. and {Eller}, P. and {Evenson}, P.~A. and {Fahey}, S. and {Fazely}, A.~R. and {Felde}, J. and {Filimonov}, K. and {Finley}, C. and {Franckowiak}, A. and {Friedman}, E. and {Fritz}, A. and {Gaisser}, T.~K. and {Gallagher}, J. and {Ganster}, E. and {Garrappa}, S. and {Gerhardt}, L. and {Ghorbani}, K. and {Glauch}, T. and {Gl{\"u}senkamp}, T. and {Goldschmidt}, A. and {Gonzalez}, J.~G. and {Grant}, D. and {Griffith}, Z. and {G{\"u}nder}, M. and {G{\"u}nd{\"u}z}, M. and {Haack}, C. and {Hallgren}, A. and {Halve}, L. and {Halzen}, F. and {Hanson}, K. and {Hebecker}, D. and {Heereman}, D. and {Helbing}, K. and {Hellauer}, R. and {Henningsen}, F. and {Hickford}, S. and {Hignight}, J. and {Hill}, G.~C. and {Hoffman}, K.~D. and {Hoffmann}, R. and {Hoinka}, T. and {Hokanson-Fasig}, B. and {Hoshina}, K. and {Huang}, F. and {Huber}, M. and {Hultqvist}, K. and {H{\"u}nnefeld}, M. and {Hussain}, R. and {In}, S. and {Iovine}, N. and {Ishihara}, A. and {Jacobi}, E. and {Japaridze}, G.~S. and {Jeong}, M. and {Jero}, K. and {Jones}, B.~J.~P. and {Kang}, W. and {Kappes}, A. and {Kappesser}, D. and {Karg}, T. and {Karl}, M. and {Karle}, A. and {Katz}, U. and {Kauer}, M. and {Keivani}, A. and {Kelley}, J.~L. and {Kheirandish}, A. and {Kim}, J. and {Kintscher}, T. and {Kiryluk}, J. and {Kittler}, T. and {Klein}, S.~R. and {Koirala}, R. and {Kolanoski}, H. and {K{\"o}pke}, L. and {Kopper}, C. and {Kopper}, S. and {Koskinen}, D.~J. and {Kowalski}, M. and {Krings}, K. and {Kr{\"u}ckl}, G. and {Kulacz}, N. and {Kunwar}, S. and {Kurahashi}, N. and {Kyriacou}, A. and {Labare}, M. and {Lanfranchi}, J.~L. and {Larson}, M.~J. and {Lauber}, F. and {Lazar}, J.~P. and {Leonard}, K. and {Leuermann}, M. and {Liu}, Q.~R. and {Lohfink}, E. and {Lozano Mariscal}, C.~J. and {Lu}, L. and {Lucarelli}, F. and {L{\"u}nemann}, J. and {Luszczak}, W. and {Madsen}, J. and {Maggi}, G. and {Mahn}, K.~B.~M. and {Makino}, Y. and {Mallot}, K. and {Mancina}, S. and {Mari{\textcommabelow s}}, I.~C. and {Maruyama}, R. and {Mase}, K. and {Maunu}, R. and {Meagher}, K. and {Medici}, M. and {Medina}, A.},
        title = "{Investigation of Two Fermi-LAT Gamma-Ray Blazars Coincident with High-energy Neutrinos Detected by IceCube}",
      journal = {\apj},
         year = 2019,
        month = aug,
       volume = {880},
       number = {2},
          eid = {103},
        pages = {103},
          doi = {10.3847/1538-4357/ab2ada},
archivePrefix = {arXiv},
       eprint = {1901.10806},
 primaryClass = {astro-ph.HE},
       adsurl = {https://ui.adsabs.harvard.edu/abs/2019ApJ...880..103G}
}

@ARTICLE{2021A&A...650A..83H,
       author = {{Hovatta}, T. and {Lindfors}, E. and {Kiehlmann}, S. and {Max-Moerbeck}, W. and {Hodges}, M. and {Liodakis}, I. and {L{\"a}hteem{\"a}ki}, A. and {Pearson}, T.~J. and {Readhead}, A.~C.~S. and {Reeves}, R.~A. and {Suutarinen}, S. and {Tammi}, J. and {Tornikoski}, M.},
        title = "{Association of IceCube neutrinos with radio sources observed at Owens Valley and Mets{\"a}hovi Radio Observatories}",
      journal = {\aap},
         year = 2021,
        month = jun,
       volume = {650},
          eid = {A83},
        pages = {A83},
          doi = {10.1051/0004-6361/202039481},
archivePrefix = {arXiv},
       eprint = {2009.10523},
 primaryClass = {astro-ph.HE},
       adsurl = {https://ui.adsabs.harvard.edu/abs/2021A&A...650A..83H}
}

@ARTICLE{2016MNRAS.455..680A,
       author = {{Agarwal}, Aditi and {Gupta}, Alok C. and {Bachev}, R. and {Strigachev}, A. and {Semkov}, E. and {Wiita}, Paul J. and {Fan}, J.~H. and {Pandey}, U.~S. and {Boeva}, S. and {Spassov}, B.},
        title = "{Multiband optical variability of the blazar S5 0716+714 in outburst state during 2014-2015}",
      journal = {\mnras},
         year = 2016,
        month = jan,
       volume = {455},
       number = {1},
        pages = {680-690},
          doi = {10.1093/mnras/stv2345},
archivePrefix = {arXiv},
       eprint = {1510.02816},
 primaryClass = {astro-ph.HE},
       adsurl = {https://ui.adsabs.harvard.edu/abs/2016MNRAS.455..680A}
}

@ARTICLE{2002A&A...390..407V,
       author = {{Villata}, M. and {Raiteri}, C.~M. and {Kurtanidze}, O.~M. and {Nikolashvili}, M.~G. and {Ibrahimov}, M.~A. and {Papadakis}, I.~E. and {Tsinganos}, K. and {Sadakane}, K. and {Okada}, N. and {Takalo}, L.~O. and {Sillanp{\"a}{\"a}}, A. and {Tosti}, G. and {Ciprini}, S. and {Frasca}, A. and {Marilli}, E. and {Robb}, R.~M. and {Noble}, J.~C. and {Jorstad}, S.~G. and {Hagen-Thorn}, V.~A. and {Larionov}, V.~M. and {Nesci}, R. and {Maesano}, M. and {Schwartz}, R.~D. and {Basler}, J. and {Gorham}, P.~W. and {Iwamatsu}, H. and {Kato}, T. and {Pullen}, C. and {Ben{\'\i}tez}, E. and {de Diego}, J.~A. and {Moilanen}, M. and {Oksanen}, A. and {Rodriguez}, D. and {Sadun}, A.~C. and {Kelly}, M. and {Carini}, M.~T. and {Miller}, H.~R. and {Catalano}, S. and {Dultzin-Hacyan}, D. and {Fan}, J.~H. and {Ishioka}, R. and {Karttunen}, H. and {Kein{\"a}nen}, P. and {Kudryavtseva}, N.~A. and {Lainela}, M. and {Lanteri}, L. and {Larionova}, E.~G. and {Matsumoto}, K. and {Mattox}, J.~R. and {Montagni}, F. and {Nucciarelli}, G. and {Ostorero}, L. and {Papamastorakis}, J. and {Pasanen}, M. and {Sobrito}, G. and {Uemura}, M.},
        title = "{The WEBT <ASTROBJ>BL Lacertae</ASTROBJ> Campaign 2000}",
      journal = {\aap},
         year = 2002,
        month = aug,
       volume = {390},
        pages = {407-421},
          doi = {10.1051/0004-6361:20020662},
archivePrefix = {arXiv},
       eprint = {astro-ph/0205479},
 primaryClass = {astro-ph},
       adsurl = {https://ui.adsabs.harvard.edu/abs/2002A&A...390..407V}
}

@ARTICLE{2005AJ....130.1418J,
       author = {{Jorstad}, Svetlana G. and {Marscher}, Alan P. and {Lister}, Matthew L. and {Stirling}, Alastair M. and {Cawthorne}, Timothy V. and {Gear}, Walter K. and {G{\'o}mez}, Jos{\'e} L. and {Stevens}, Jason A. and {Smith}, Paul S. and {Forster}, James R. and {Robson}, E. Ian},
        title = "{Polarimetric Observations of 15 Active Galactic Nuclei at High Frequencies: Jet Kinematics from Bimonthly Monitoring with the Very Long Baseline Array}",
      journal = {\aj},
         year = 2005,
        month = oct,
       volume = {130},
       number = {4},
        pages = {1418-1465},
          doi = {10.1086/444593},
archivePrefix = {arXiv},
       eprint = {astro-ph/0502501},
 primaryClass = {astro-ph},
       adsurl = {https://ui.adsabs.harvard.edu/abs/2005AJ....130.1418J}
}

@ARTICLE{2014ApJ...780...87M,
       author = {{Marscher}, Alan P.},
        title = "{Turbulent, Extreme Multi-zone Model for Simulating Flux and Polarization Variability in Blazars}",
      journal = {\apj},
         year = 2014,
        month = jan,
       volume = {780},
       number = {1},
          eid = {87},
        pages = {87},
          doi = {10.1088/0004-637X/780/1/87},
archivePrefix = {arXiv},
       eprint = {1311.7665},
 primaryClass = {astro-ph.HE},
       adsurl = {https://ui.adsabs.harvard.edu/abs/2014ApJ...780...87M}
}

@ARTICLE{2020ApJ...902...41W,
       author = {{Wang}, Yi-Fan and {Jiang}, Yun-Guo},
        title = "{A Comprehensive Study on the Variation Phenomena of AO 0235+164}",
      journal = {\apj},
         year = 2020,
        month = oct,
       volume = {902},
       number = {1},
          eid = {41},
        pages = {41},
          doi = {10.3847/1538-4357/abb36c},
archivePrefix = {arXiv},
       eprint = {2009.00879},
 primaryClass = {astro-ph.HE},
       adsurl = {https://ui.adsabs.harvard.edu/abs/2020ApJ...902...41W}
}

@ARTICLE{2007AJ....133.1599W,
       author = {{Wu}, Jianghua and {Zhou}, Xu and {Ma}, Jun and {Wu}, Zhenyu and {Jiang}, Zhaoji and {Chen}, Jiansheng},
        title = "{Optical Monitoring of BL Lacertae Object S5 0716+714 with a Novel Multipeak Interference Filter}",
      journal = {\aj},
         year = 2007,
        month = apr,
       volume = {133},
       number = {4},
        pages = {1599-1608},
          doi = {10.1086/511773},
archivePrefix = {arXiv},
       eprint = {astro-ph/0612142},
 primaryClass = {astro-ph},
       adsurl = {https://ui.adsabs.harvard.edu/abs/2007AJ....133.1599W}
}

@ARTICLE{2026ApJ..1001...32X,
       author = {{Xia}, Hanxiao and {Wang}, Ziming and {Wu}, Jianghua and {Fang}, Yue and {Du}, Shiyu},
        title = "{Multiple Components and Spectral Evolution of BL Lacertae as Revealed by Multiwavelength Variability and SED Modeling}",
      journal = {\apj},
         year = 2026,
        month = apr,
       volume = {1001},
       number = {1},
          eid = {32},
        pages = {32},
          doi = {10.3847/1538-4357/ae4dde},
archivePrefix = {arXiv},
       eprint = {2511.13317},
 primaryClass = {astro-ph.HE},
       adsurl = {https://ui.adsabs.harvard.edu/abs/2026ApJ..1001...32X}
}

@ARTICLE{2021BlgAJ..34...79B,
       author = {{Bachev}, R. and {Strigachev}, A. and {Kurtenkov}, A. and {Spassov}, B. and {Nikolov}, Y. and {Boeva}, S. and {Semkov}, E.},
        title = "{Optical follow-up of TXS 0506+056 after the neutrino detection}",
      journal = {Bulgarian Astronomical Journal},
         year = 2021,
        month = feb,
       volume = {34},
        pages = {79},
       adsurl = {https://ui.adsabs.harvard.edu/abs/2021BlgAJ..34...79B}
}

@ARTICLE{2021PASJ...73...25M,
       author = {{Morokuma}, Tomoki and {Utsumi}, Yousuke and {Ohta}, Kouji and {Yamanaka}, Masayuki and {Kawabata}, Koji S. and {Inoue}, Yoshiyuki and {Tanaka}, Masaomi and {Yoshida}, Michitoshi and {Itoh}, Ryosuke and {Sasada}, Mahito and et al.},
        title = "{Follow-up observations for IceCube-170922A: Detection of rapid near-infrared variability and intensive monitoring of TXS 0506+056}",
      journal = {\pasj},
         year = 2021,
        month = feb,
       volume = {73},
       number = {1},
        pages = {25-43},
          doi = {10.1093/pasj/psaa110},
archivePrefix = {arXiv},
       eprint = {2011.04957},
 primaryClass = {astro-ph.HE},
       adsurl = {https://ui.adsabs.harvard.edu/abs/2021PASJ...73...25M}
}

@ARTICLE{2019A&A...630A.103B,
       author = {{Britzen}, S. and {Fendt}, C. and {B{\"o}ttcher}, M. and {Zaja{\v{c}}ek}, M. and {Jaron}, F. and {Pashchenko}, I.~N. and {Araudo}, A. and {Karas}, V. and {Kurtanidze}, O.},
        title = "{A cosmic collider: Was the IceCube neutrino generated in a precessing jet-jet interaction in TXS 0506+056?}",
      journal = {\aap},
         year = 2019,
        month = oct,
       volume = {630},
          eid = {A103},
        pages = {A103},
          doi = {10.1051/0004-6361/201935422},
       adsurl = {https://ui.adsabs.harvard.edu/abs/2019A&A...630A.103B}
}

@INPROCEEDINGS{2023mgm..conf.1467L,
       author = {{Lipunov}, V.~M. and {Zhirkov}, K. and {Kornilov}, V.~G. and {Gorbovskoy}, E. and {Budnev}, N.~M. and {Buckley}, D.~A.~H. and {Rebolo}, R. and {Serra}, M. and {Podesta}, R. and {Francile}, C. and et al.},
        title = "{MASTER optical observations of the blazar TXS0506+056 during the IC170922A}",
    booktitle = {The Sixteenth Marcel Grossmann Meeting. On Recent Developments in Theoretical and Experimental General Relativity, Astrophysics, and Relativistic Field Theories},
         year = 2023,
       editor = {{Ruffino}, Remo and {Vereshchagin}, Gregory},
        month = jul,
        pages = {1467-1473},
          doi = {10.1142/9789811269776_0117},
       adsurl = {https://ui.adsabs.harvard.edu/abs/2023mgm..conf.1467L}
}

@ARTICLE{2018RNAAS...2..130G,
       author = {{Gab{\'a}nyi}, Krisztina {\'E}va and {Mo{\'o}r}, Attila and {Frey}, S{\'a}ndor},
        title = "{Mid-infrared Variability of the Neutrino Source Blazar TXS 0506+056}",
      journal = {Research Notes of the American Astronomical Society},
         year = 2018,
        month = jul,
       volume = {2},
       number = {3},
          eid = {130},
        pages = {130},
          doi = {10.3847/2515-5172/aad49f},
archivePrefix = {arXiv},
       eprint = {1807.07462},
 primaryClass = {astro-ph.GA},
       adsurl = {https://ui.adsabs.harvard.edu/abs/2018RNAAS...2..130G}
}

@ARTICLE{2019MNRAS.484L.104P,
       author = {{Padovani}, P. and {Oikonomou}, F. and {Petropoulou}, M. and {Giommi}, P. and {Resconi}, E.},
        title = "{TXS 0506+056, the first cosmic neutrino source, is not a BL Lac}",
      journal = {\mnras},
         year = 2019,
        month = mar,
       volume = {484},
       number = {1},
        pages = {L104-L108},
          doi = {10.1093/mnrasl/slz011},
archivePrefix = {arXiv},
       eprint = {1901.06998},
 primaryClass = {astro-ph.HE},
       adsurl = {https://ui.adsabs.harvard.edu/abs/2019MNRAS.484L.104P}
}

@ARTICLE{2017ApJ...844..107I,
       author = {{Isler}, Jedidah C. and {Urry}, C.~M. and {Coppi}, P. and {Bailyn}, C. and {Brady}, M. and {MacPherson}, E. and {Buxton}, M. and {Hasan}, I.},
        title = "{A Consolidated Framework of the Color Variability in Blazars: Long-term Optical/Near-infrared Observations of 3C 279}",
      journal = {\apj},
         year = 2017,
        month = aug,
       volume = {844},
       number = {2},
          eid = {107},
        pages = {107},
          doi = {10.3847/1538-4357/aa79fc},
archivePrefix = {arXiv},
       eprint = {1706.09891},
 primaryClass = {astro-ph.HE},
       adsurl = {https://ui.adsabs.harvard.edu/abs/2017ApJ...844..107I}
}

@ARTICLE{2012MNRAS.425.3002G,
       author = {{Gaur}, Haritma and {Gupta}, Alok C. and {Strigachev}, A. and {Bachev}, R. and {Semkov}, E. and {Wiita}, Paul J. and {Peneva}, S. and {Boeva}, S. and {Slavcheva-Mihova}, L. and {Mihov}, B. and et al.},
        title = "{Optical flux and spectral variability of blazars}",
      journal = {\mnras},
         year = 2012,
        month = oct,
       volume = {425},
       number = {4},
        pages = {3002-3023},
          doi = {10.1111/j.1365-2966.2012.21583.x},
archivePrefix = {arXiv},
       eprint = {1207.5943},
 primaryClass = {astro-ph.HE},
       adsurl = {https://ui.adsabs.harvard.edu/abs/2012MNRAS.425.3002G}
}

@ARTICLE{2006A&A...450...39G,
       author = {{Gu}, M.~F. and {Lee}, C.-U. and {Pak}, S. and {Yim}, H.~S. and {Fletcher}, A.~B.},
        title = "{Multi-colour optical monitoring of eight red blazars}",
      journal = {\aap},
         year = 2006,
        month = apr,
       volume = {450},
       number = {1},
        pages = {39-51},
          doi = {10.1051/0004-6361:20054271},
archivePrefix = {arXiv},
       eprint = {astro-ph/0602180},
 primaryClass = {astro-ph},
       adsurl = {https://ui.adsabs.harvard.edu/abs/2006A&A...450...39G}
}

@ARTICLE{2021APh...12902577B,
       author = {{Butuzova}, M.~S.},
        title = "{A geometrical interpretation for the properties of multiband optical variability of the blazar S5 0716+714}",
      journal = {Astroparticle Physics},
         year = 2021,
        month = may,
       volume = {129},
          eid = {102577},
        pages = {102577},
          doi = {10.1016/j.astropartphys.2021.102577},
archivePrefix = {arXiv},
       eprint = {2005.08161},
 primaryClass = {astro-ph.GA},
       adsurl = {https://ui.adsabs.harvard.edu/abs/2021APh...12902577B}
}

@ARTICLE{2015MNRAS.454..353R,
       author = {{Raiteri}, C.~M. and {Stamerra}, A. and {Villata}, M. and {Larionov}, V.~M. and {Acosta-Pulido}, J.~A. and {Ar{\'e}valo}, M.~J. and {Arkharov}, A.~A. and {Bachev}, R. and {Ben{\'\i}tez}, E. and {Bozhilov}, V.~V. and {Borman}, G.~A. and {Buemi}, C.~S. and {Calcidese}, P. and {Carnerero}, M.~I. and {Carosati}, D. and {Chigladze}, R.~A. and {Damljanovic}, G. and {Di Paola}, A. and {Doroshenko}, V.~T. and {Efimova}, N.~V. and {Ehgamberdiev}, Sh. A. and {Giroletti}, M. and {Gonz{\'a}lez-Morales}, P.~A. and {Grinon-Marin}, A.~B. and {Grishina}, T.~S. and {Hiriart}, D. and {Ibryamov}, S. and {Klimanov}, S.~A. and {Kopatskaya}, E.~N. and {Kurtanidze}, O.~M. and {Kurtanidze}, S.~O. and {Kurtenkov}, A.~A. and {Larionova}, L.~V. and {Larionova}, E.~G. and {L{\'a}zaro}, C. and {L{\"a}hteenm{\"a}ki}, A. and {Leto}, P. and {Markovic}, G. and {Mirzaqulov}, D.~O. and {Mokrushina}, A.~A. and {Morozova}, D.~A. and {M{\'u}jica}, R. and {Nazarov}, S.~V. and {Nikolashvili}, M.~G. and {Ohlert}, J.~M. and {Ovcharov}, E.~P. and {Paiano}, S. and {Pastor Yabar}, A. and {Prandini}, E. and {Ramakrishnan}, V. and {Sadun}, A.~C. and {Semkov}, E. and {Sigua}, L.~A. and {Strigachev}, A. and {Tammi}, J. and {Tornikoski}, M. and {Trigilio}, C. and {Troitskaya}, Yu. V. and {Troitsky}, I.~S. and {Umana}, G. and {Velasco}, S. and {Vince}, O.},
        title = "{The WEBT campaign on the BL Lac object PG 1553+113 in 2013. An analysis of the enigmatic synchrotron emission}",
      journal = {\mnras},
         year = 2015,
        month = nov,
       volume = {454},
       number = {1},
        pages = {353-367},
          doi = {10.1093/mnras/stv1884},
archivePrefix = {arXiv},
       eprint = {1509.02706},
 primaryClass = {astro-ph.HE},
       adsurl = {https://ui.adsabs.harvard.edu/abs/2015MNRAS.454..353R}
}

@ARTICLE{2011MNRAS.418.1640W,
       author = {{Wu}, Jianghua and {Zhou}, Xu and {Ma}, Jun and {Jiang}, Zhaoji},
        title = "{Optical variability and colour behaviour of 3C 345}",
      journal = {\mnras},
         year = 2011,
        month = dec,
       volume = {418},
       number = {3},
        pages = {1640-1648},
          doi = {10.1111/j.1365-2966.2011.19565.x},
archivePrefix = {arXiv},
       eprint = {1108.1020},
 primaryClass = {astro-ph.CO},
       adsurl = {https://ui.adsabs.harvard.edu/abs/2011MNRAS.418.1640W}
}

@ARTICLE{2009ApJ...694..174B,
       author = {{B{\"o}ttcher}, M. and {Fultz}, K. and {Aller}, H.~D. and {Aller}, M.~F. and {Apodaca}, J. and {Arkharov}, A.~A. and {Bach}, U. and {Bachev}, R. and {Berdyugin}, A. and {Buemi}, C. and {Calcidese}, P. and {Carosati}, D. and {Charlot}, P. and {Ciprini}, S. and {di Paola}, A. and {Dolci}, M. and {Efimova}, N.~V. and {Scurrats}, E. Forn{\'e} and {Frasca}, A. and {Gupta}, A.~C. and {Hagen-Thorn}, V.~A. and {Heidt}, J. and {Hiriart}, D. and {Konstantinova}, T.~S. and {Kopatskaya}, E.~N. and {L{\"a}hteenm{\"a}ki}, A. and {Lanteri}, L. and {Larionov}, V.~M. and {LeCampion}, J.-F. and {Leto}, P. and {Lindfors}, E. and {Mihov}, B. and {Marilli}, B. and {Nieppola}, E. and {Nilsson}, K. and {Ohlert}, J.~M. and {Ovcharov}, E. and {P{\"a}{\"a}kk{\"o}nen}, P. and {Pasanen}, M. and {Ragozzine}, B. and {Raiteri}, C.~M. and {Ros}, J.~A. and {Sadun}, A. and {Sanchez}, A. and {Semkov}, E. and {Sorcia}, M. and {Strigachev}, A. and {Takalo}, L. and {Tornikoski}, M. and {Trigilio}, C. and {Umana}, G. and {Valcheva}, A. and {Villata}, M. and {Volvach}, A. and {Wu}, J.-H. and {Zhou}, X.},
        title = "{The Whole Earth Blazar Telescope Campaign on the Intermediate BL Lac Object 3C 66A in 2007-2008}",
      journal = {\apj},
         year = 2009,
        month = mar,
       volume = {694},
       number = {1},
        pages = {174-182},
          doi = {10.1088/0004-637X/694/1/174},
archivePrefix = {arXiv},
       eprint = {0811.0501},
 primaryClass = {astro-ph},
       adsurl = {https://ui.adsabs.harvard.edu/abs/2009ApJ...694..174B}
}

@ARTICLE{2012ApJ...756...13B,
       author = {{Bonning}, Erin and {Urry}, C. Megan and {Bailyn}, Charles and {Buxton}, Michelle and {Chatterjee}, Ritaban and {Coppi}, Paolo and {Fossati}, Giovanni and {Isler}, Jedidah and {Maraschi}, Laura},
        title = "{SMARTS Optical and Infrared Monitoring of 12 Gamma-Ray Bright Blazars}",
      journal = {\apj},
         year = 2012,
        month = sep,
       volume = {756},
       number = {1},
          eid = {13},
        pages = {13},
          doi = {10.1088/0004-637X/756/1/13},
archivePrefix = {arXiv},
       eprint = {1201.4380},
 primaryClass = {astro-ph.HE},
       adsurl = {https://ui.adsabs.harvard.edu/abs/2012ApJ...756...13B}
}

@ARTICLE{2005AJ....129.1818W,
       author = {{Wu}, Jianghua and {Peng}, Bo and {Zhou}, Xu and {Ma}, Jun and {Jiang}, Zhaoji and {Chen}, Jiansheng},
        title = "{Optical Monitoring of BL Lacertae Object S5 0716+714 with High Temporal Resolution}",
      journal = {\aj},
         year = 2005,
        month = apr,
       volume = {129},
       number = {4},
        pages = {1818-1826},
          doi = {10.1086/428599},
archivePrefix = {arXiv},
       eprint = {astro-ph/0501184},
 primaryClass = {astro-ph},
       adsurl = {https://ui.adsabs.harvard.edu/abs/2005AJ....129.1818W}
}

@ARTICLE{1996MNRAS.281..425M,
       author = {{Marcha}, M.~J.~M. and {Browne}, I.~W.~A. and {Impey}, C.~D. and {Smith}, P.~S.},
        title = "{Optical spectroscopy and polarization of a new sample of optically bright flat radio spectrum sources}",
      journal = {\mnras},
         year = 1996,
        month = jul,
       volume = {281},
       number = {2},
        pages = {425-448},
          doi = {10.1093/mnras/281.2.425},
       adsurl = {https://ui.adsabs.harvard.edu/abs/1996MNRAS.281..425M}
}

@ARTICLE{2025PhRvD.112h3016W,
       author = {{Wang}, Zhen-Jie and {Liu}, Ruo-Yu and {Wang}, Xiang-Yu},
        title = "{Stochastic dissipation model for the steady state neutrino and multiwavelength emissions of TXS 0506+056}",
      journal = {\prd},
         year = 2025,
        month = oct,
       volume = {112},
       number = {8},
          eid = {083016},
        pages = {083016},
          doi = {10.1103/w7mw-gghk},
archivePrefix = {arXiv},
       eprint = {2509.14587},
 primaryClass = {astro-ph.HE},
       adsurl = {https://ui.adsabs.harvard.edu/abs/2025PhRvD.112h3016W}
}

@ARTICLE{2024ApJ...962..142W,
       author = {{Wang}, Zhen-Jie and {Liu}, Ruo-Yu and {Wang}, Ze-Rui and {Wang}, Junfeng},
        title = "{A Unified Model for Multiepoch Neutrino Events and Broadband Spectral Energy Distribution of TXS 0506+056}",
      journal = {\apj},
         year = 2024,
        month = feb,
       volume = {962},
       number = {2},
          eid = {142},
        pages = {142},
          doi = {10.3847/1538-4357/ad1bca},
archivePrefix = {arXiv},
       eprint = {2401.06304},
 primaryClass = {astro-ph.HE},
       adsurl = {https://ui.adsabs.harvard.edu/abs/2024ApJ...962..142W}
}

@ARTICLE{2021ApJ...906...51X,
       author = {{Xue}, Rui and {Liu}, Ruo-Yu and {Wang}, Ze-Rui and {Ding}, Nan and {Wang}, Xiang-Yu},
        title = "{A Two-zone Blazar Radiation Model for ``Orphan'' Neutrino Flares}",
      journal = {\apj},
         year = 2021,
        month = jan,
       volume = {906},
       number = {1},
          eid = {51},
        pages = {51},
          doi = {10.3847/1538-4357/abc886},
archivePrefix = {arXiv},
       eprint = {2011.03681},
 primaryClass = {astro-ph.HE},
       adsurl = {https://ui.adsabs.harvard.edu/abs/2021ApJ...906...51X}
}

@ARTICLE{1996SSRv...75..341S,
       author = {{Stecker}, F.~W. and {Salamon}, M.~H.},
        title = "{High Energy Neutrinos from Quasars}",
      journal = {\ssr},
         year = 1996,
        month = jan,
       volume = {75},
       number = {1-2},
        pages = {341-355},
          doi = {10.1007/BF00195044},
archivePrefix = {arXiv},
       eprint = {astro-ph/9501064},
 primaryClass = {astro-ph},
       adsurl = {https://ui.adsabs.harvard.edu/abs/1996SSRv...75..341S}
}

@ARTICLE{1995APh.....3..295M,
       author = {{Mannheim}, Karl},
        title = "{High-energy neutrinos from extragalactic jets}",
      journal = {Astroparticle Physics},
         year = 1995,
        month = may,
       volume = {3},
       number = {3},
        pages = {295-302},
          doi = {10.1016/0927-6505(94)00044-4},
       adsurl = {https://ui.adsabs.harvard.edu/abs/1995APh.....3..295M}
}

@ARTICLE{1998A&A...333..452K,
       author = {{Kirk}, J.~G. and {Rieger}, F.~M. and {Mastichiadis}, A.},
        title = "{Particle acceleration and synchrotron emission in blazar jets}",
      journal = {\aap},
         year = 1998,
        month = may,
       volume = {333},
        pages = {452-458},
          doi = {10.48550/arXiv.astro-ph/9801265},
archivePrefix = {arXiv},
       eprint = {astro-ph/9801265},
 primaryClass = {astro-ph},
       adsurl = {https://ui.adsabs.harvard.edu/abs/1998A&A...333..452K}
}

@ARTICLE{1985ApJ...298..114M,
       author = {{Marscher}, A.~P. and {Gear}, W.~K.},
        title = "{Models for high-frequency radio outbursts in extragalactic sources, with application to the early 1983 millimeter-to-infrared flare of 3C 273.}",
      journal = {\apj},
         year = 1985,
        month = nov,
       volume = {298},
        pages = {114-127},
          doi = {10.1086/163592},
       adsurl = {https://ui.adsabs.harvard.edu/abs/1985ApJ...298..114M}
}

@ARTICLE{2018Sci...361.1378I,
       author = {{IceCube Collaboration} and {Aartsen}, M.~G. and {Ackermann}, M. and {Adams}, J. and {Aguilar}, J.~A. and {Ahlers}, M. and {Ahrens}, M. and {Al Samarai}, I. and {Altmann}, D. and {Andeen}, K. and {Anderson}, T. and {Ansseau}, I. and {Anton}, G. and {Arg{\"u}elles}, C. and {Auffenberg}, J. and {Axani}, S. and {Bagherpour}, H. and {Bai}, X. and {Barron}, J.~P. and {Barwick}, S.~W. and {Baum}, V. and {Bay}, R. and {Beatty}, J.~J. and {Becker Tjus}, J. and {Becker}, K.-H. and {BenZvi}, S. and {Berley}, D. and {Bernardini}, E. and {Besson}, D.~Z. and {Binder}, G. and {Bindig}, D. and {Blaufuss}, E. and {Blot}, S. and {Bohm}, C. and {B{\"o}rner}, M. and {Bos}, F. and {B{\"o}ser}, S. and {Botner}, O. and {Bourbeau}, E. and {Bourbeau}, J. and {Bradascio}, F. and {Braun}, J. and {Brenzke}, M. and {Bretz}, H.-P. and {Bron}, S. and {Brostean-Kaiser}, J. and {Burgman}, A. and {Busse}, R.~S. and {Carver}, T. and {Cheung}, E. and {Chirkin}, D. and {Christov}, A. and {Clark}, K. and {Classen}, L. and {Coenders}, S. and {Collin}, G.~H. and {Conrad}, J.~M. and {Coppin}, P. and {Correa}, P. and {Cowen}, D.~F. and {Cross}, R. and {Dave}, P. and {Day}, M. and {de Andr{\'e}}, J.~P.~A.~M. and {De Clercq}, C. and {DeLaunay}, J.~J. and {Dembinski}, H. and {De Ridder}, S. and {Desiati}, P. and {de Vries}, K.~D. and {de Wasseige}, G. and {de With}, M. and {DeYoung}, T. and {D{\'\i}az-V{\'e}lez}, J.~C. and {di Lorenzo}, V. and {Dujmovic}, H. and {Dumm}, J.~P. and {Dunkman}, M. and {Dvorak}, E. and {Eberhardt}, B. and {Ehrhardt}, T. and {Eichmann}, B. and {Eller}, P. and {Evenson}, P.~A. and {Fahey}, S. and {Fazely}, A.~R. and {Felde}, J. and {Filimonov}, K. and {Finley}, C. and {Flis}, S. and {Franckowiak}, A. and {Friedman}, E. and {Fritz}, A. and {Gaisser}, T.~K. and {Gallagher}, J. and {Gerhardt}, L. and {Ghorbani}, K. and {Glauch}, T. and {Gl{\"u}senkamp}, T. and {Goldschmidt}, A. and {Gonzalez}, J.~G. and {Grant}, D. and {Griffith}, Z. and {Haack}, C. and {Hallgren}, A. and {Halzen}, F. and {Hanson}, K. and {Hebecker}, D. and {Heereman}, D. and {Helbing}, K. and {Hellauer}, R. and {Hickford}, S. and {Hignight}, J. and {Hill}, G.~C. and {Hoffman}, K.~D. and {Hoffmann}, R. and {Hoinka}, T. and {Hokanson-Fasig}, B. and {Hoshina}, K. and {Huang}, F. and {Huber}, M. and {Hultqvist}, K. and {H{\"u}nnefeld}, M. and {Hussain}, R. and {In}, S. and {Iovine}, N. and {Ishihara}, A. and {Jacobi}, E. and {Japaridze}, G.~S. and {Jeong}, M. and {Jero}, K. and {Jones}, B.~J.~P. and {Kalaczynski}, P. and {Kang}, W. and {Kappes}, A. and {Kappesser}, D. and {Karg}, T. and {Karle}, A. and {Katz}, U. and {Kauer}, M. and {Keivani}, A. and {Kelley}, J.~L. and {Kheirandish}, A. and {Kim}, J. and {Kim}, M. and {Kintscher}, T. and {Kiryluk}, J. and {Kittler}, T. and {Klein}, S.~R. and {Koirala}, R. and {Kolanoski}, H. and {K{\"o}pke}, L. and {Kopper}, C. and {Kopper}, S. and {Koschinsky}, J.~P. and {Koskinen}, D.~J. and {Kowalski}, M. and {Krings}, K. and {Kroll}, M. and {Kr{\"u}ckl}, G. and {Kunwar}, S. and {Kurahashi}, N. and {Kuwabara}, T. and {Kyriacou}, A. and {Labare}, M. and {Lanfranchi}, J.~L. and {Larson}, M.~J. and {Lauber}, F. and {Leonard}, K. and {Lesiak-Bzdak}, M. and {Leuermann}, M. and {Liu}, Q.~R. and {Lozano Mariscal}, C.~J. and {Lu}, L. and {L{\"u}nemann}, J. and {Luszczak}, W. and {Madsen}, J. and {Maggi}, G. and {Mahn}, K.~B.~M. and {Mancina}, S. and {Maruyama}, R. and {Mase}, K. and {Maunu}, R. and {Meagher}, K. and {Medici}, M. and {Meier}, M. and {Menne}, T. and {Merino}, G. and {Meures}, T. and {Miarecki}, S. and {Micallef}, J. and {Moment{\'e}}, G. and {Montaruli}, T. and {Moore}, R.~W. and {Morse}, R. and {Moulai}, M. and {Nahnhauer}, R. and {Nakarmi}, P. and {Naumann}, U. and {Neer}, G.},
       title = "{Multimessenger observations of a flaring blazar coincident with high-energy neutrino IceCube-170922A}",
      journal = {Science},
         year = 2018,
        month = jul,
       volume = {361},
       number = {6398},
          eid = {eaat1378},
        pages = {eaat1378},
          doi = {10.1126/science.aat1378},
archivePrefix = {arXiv},
       eprint = {1807.08816},
 primaryClass = {astro-ph.HE},
       adsurl = {https://ui.adsabs.harvard.edu/abs/2018Sci...361.1378I}
}

@ARTICLE{2018Sci...361..147I,
       author = {{IceCube Collaboration} and {Aartsen}, M.~G. and {Ackermann}, M. and {Adams}, J. and {Aguilar}, J.~A. and {Ahlers}, M. and {Ahrens}, M. and {Samarai}, I. Al and {Altmann}, D. and {Andeen}, K. and {Anderson}, T. and {Ansseau}, I. and {Anton}, G. and {Arg{\"u}elles}, C. and {Arsioli}, B. and {Auffenberg}, J. and {Axani}, S. and {Bagherpour}, H. and {Bai}, X. and {Barron}, J.~P. and {Barwick}, S.~W. and {Baum}, V. and {Bay}, R. and {Beatty}, J.~J. and {Becker Tjus}, J. and {Becker}, K.-H. and {BenZvi}, S. and {Berley}, D. and {Bernardini}, E. and {Besson}, D.~Z. and {Binder}, G. and {Bindig}, D. and {Blaufuss}, E. and {Blot}, S. and {Bohm}, C. and {B{\"o}rner}, M. and {Bos}, F. and {B{\"o}ser}, S. and {Botner}, O. and {Bourbeau}, E. and {Bourbeau}, J. and {Bradascio}, F. and {Braun}, J. and {Brenzke}, M. and {Bretz}, H.-P. and {Bron}, S. and {Brostean-Kaiser}, J. and {Burgman}, A. and {Busse}, R.~S. and {Carver}, T. and {Cheung}, E. and {Chirkin}, D. and {Christov}, A. and {Clark}, K. and {Classen}, L. and {Coenders}, S. and {Collin}, G.~H. and {Conrad}, J.~M. and {Coppin}, P. and {Correa}, P. and {Cowen}, D.~F. and {Cross}, R. and {Dave}, P. and {Day}, M. and {de Andr{\'e}}, J.~P.~A.~M. and {De Clercq}, C. and {DeLaunay}, J.~J. and {Dembinski}, H. and {DeRidder}, S. and {Desiati}, P. and {de Vries}, K.~D. and {de Wasseige}, G. and {de With}, M. and {DeYoung}, T. and {D{\'\i}az-V{\'e}lez}, J.~C. and {di Lorenzo}, V. and {Dujmovic}, H. and {Dumm}, J.~P. and {Dunkman}, M. and {Dvorak}, E. and {Eberhardt}, B. and {Ehrhardt}, T. and {Eichmann}, B. and {Eller}, P. and {Evenson}, P.~A. and {Fahey}, S. and {Fazely}, A.~R. and {Felde}, J. and {Filimonov}, K. and {Finley}, C. and {Flis}, S. and {Franckowiak}, A. and {Friedman}, E. and {Fritz}, A. and {Gaisser}, T.~K. and {Gallagher}, J. and {Gerhardt}, L. and {Ghorbani}, K. and {Giommi}, P. and {Glauch}, T. and {Gl{\"u}senkamp}, T. and {Goldschmidt}, A. and {Gonzalez}, J.~G. and {Grant}, D. and {Griffith}, Z. and {Haack}, C. and {Hallgren}, A. and {Halzen}, F. and {Hanson}, K. and {Hebecker}, D. and {Heereman}, D. and {Helbing}, K. and {Hellauer}, R. and {Hickford}, S. and {Hignight}, J. and {Hill}, G.~C. and {Hoffman}, K.~D. and {Hoffmann}, R. and {Hoinka}, T. and {Hokanson-Fasig}, B. and {Hoshina}, K. and {Huang}, F. and {Huber}, M. and {Hultqvist}, K. and {H{\"u}nnefeld}, M. and {Hussain}, R. and {In}, S. and {Iovine}, N. and {Ishihara}, A. and {Jacobi}, E. and {Japaridze}, G.~S. and {Jeong}, M. and {Jero}, K. and {Jones}, B.~J.~P. and {Kalaczynski}, P. and {Kang}, W. and {Kappes}, A. and {Kappesser}, D. and {Karg}, T. and {Karle}, A. and {Katz}, U. and {Kauer}, M. and {Keivani}, A. and {Kelley}, J.~L. and {Kheirandish}, A. and {Kim}, J. and {Kim}, M. and {Kintscher}, T. and {Kiryluk}, J. and {Kittler}, T. and {Klein}, S.~R. and {Koirala}, R. and {Kolanoski}, H. and {K{\"o}pke}, L. and {Kopper}, C. and {Kopper}, S. and {Koschinsky}, J.~P. and {Koskinen}, D.~J. and {Kowalski}, M. and {Krammer}, B. and {Krings}, K. and {Kroll}, M. and {Kr{\"u}ckl}, G. and {Kunwar}, S. and {Kurahashi}, N. and {Kuwabara}, T. and {Kyriacou}, A. and {Labare}, M. and {Lanfranchi}, J.~L. and {Larson}, M.~J. and {Lauber}, F. and {Leonard}, K. and {Lesiak-Bzdak}, M. and {Leuermann}, M. and {Liu}, Q.~R. and {Lozano Mariscal}, C.~J. and {Lu}, L. and {L{\"u}nemann}, J. and {Luszczak}, W. and {Madsen}, J. and {Maggi}, G. and {Mahn}, K.~B.~M. and {Mancina}, S. and {Maruyama}, R. and {Mase}, K. and {Maunu}, R. and {Meagher}, K. and {Medici}, M. and {Meier}, M. and {Menne}, T. and {Merino}, G. and {Meures}, T. and {Miarecki}, S. and {Micallef}, J. and {Moment{\'e}}, G. and {Montaruli}, T. and {Moore}, R.~W. and {Morse}, R. and {Moulai}, M. and {Nahnhauer}, R.},
        title = "{Neutrino emission from the direction of the blazar TXS 0506+056 prior to the IceCube-170922A alert}",
      journal = {Science},
         year = 2018,
        month = jul,
       volume = {361},
       number = {6398},
        pages = {147-151},
          doi = {10.1126/science.aat2890},
archivePrefix = {arXiv},
       eprint = {1807.08794},
 primaryClass = {astro-ph.HE},
       adsurl = {https://ui.adsabs.harvard.edu/abs/2018Sci...361..147I}
}

@ARTICLE{2018ApJ...854L..32P,
       author = {{Paiano}, Simona and {Falomo}, Renato and {Treves}, Aldo and {Scarpa}, Riccardo},
        title = "{The Redshift of the BL Lac Object TXS 0506+056}",
      journal = {\apjl},
         year = 2018,
        month = feb,
       volume = {854},
       number = {2},
          eid = {L32},
        pages = {L32},
          doi = {10.3847/2041-8213/aaad5e},
archivePrefix = {arXiv},
       eprint = {1802.01939},
 primaryClass = {astro-ph.GA},
       adsurl = {https://ui.adsabs.harvard.edu/abs/2018ApJ...854L..32P}
}

@ARTICLE{2004A&A...422..505G,
       author = {{Gupta}, A.~C. and {Banerjee}, D.~P.~K. and {Ashok}, N.~M. and {Joshi}, U.~C.},
        title = "{Near infrared intraday variability of Mrk 421}",
      journal = {\aap},
         year = 2004,
        month = aug,
       volume = {422},
        pages = {505-508},
          doi = {10.1051/0004-6361:20040306},
archivePrefix = {arXiv},
       eprint = {astro-ph/0405186},
 primaryClass = {astro-ph},
       adsurl = {https://ui.adsabs.harvard.edu/abs/2004A&A...422..505G}
}

@ARTICLE{1995ARA&A..33..163W,
       author = {{Wagner}, S.~J. and {Witzel}, A.},
        title = "{Intraday Variability In Quasars and BL Lac Objects}",
      journal = {\araa},
         year = 1995,
        month = jan,
       volume = {33},
        pages = {163-198},
          doi = {10.1146/annurev.aa.33.090195.001115},
       adsurl = {https://ui.adsabs.harvard.edu/abs/1995ARA&A..33..163W}
}

@ARTICLE{2003APh....18..593M,
       author = {{M{\"u}cke}, A. and {Protheroe}, R.~J. and {Engel}, R. and {Rachen}, J.~P. and {Stanev}, T.},
        title = "{BL Lac objects in the synchrotron proton blazar model}",
      journal = {Astroparticle Physics},
         year = 2003,
        month = mar,
       volume = {18},
       number = {6},
        pages = {593-613},
          doi = {10.1016/S0927-6505(02)00185-8},
archivePrefix = {arXiv},
       eprint = {astro-ph/0206164},
 primaryClass = {astro-ph},
       adsurl = {https://ui.adsabs.harvard.edu/abs/2003APh....18..593M}
}

@ARTICLE{2007Ap&SS.307...69B,
       author = {{B{\"o}ttcher}, Markus},
        title = "{Astrophysical Jets of Blazars and Microquasars}",
      journal = {\apss},
         year = 2007,
        month = jan,
       volume = {307},
       number = {1-3},
        pages = {69-75},
          doi = {10.1007/s10509-006-9213-x},
       adsurl = {https://ui.adsabs.harvard.edu/abs/2007Ap&SS.307...69B}
}

@ARTICLE{1998MNRAS.299..433F,
       author = {{Fossati}, G. and {Maraschi}, L. and {Celotti}, A. and {Comastri}, A. and {Ghisellini}, G.},
        title = "{A unifying view of the spectral energy distributions of blazars}",
      journal = {\mnras},
         year = 1998,
        month = sep,
       volume = {299},
       number = {2},
        pages = {433-448},
          doi = {10.1046/j.1365-8711.1998.01828.x},
archivePrefix = {arXiv},
       eprint = {astro-ph/9804103},
 primaryClass = {astro-ph},
       adsurl = {https://ui.adsabs.harvard.edu/abs/1998MNRAS.299..433F}
}

@ARTICLE{1995PASP..107..803U,
       author = {{Urry}, C. Megan and {Padovani}, Paolo},
        title = "{Unified Schemes for Radio-Loud Active Galactic Nuclei}",
      journal = {\pasp},
         year = 1995,
        month = sep,
       volume = {107},
        pages = {803},
          doi = {10.1086/133630},
archivePrefix = {arXiv},
       eprint = {astro-ph/9506063},
 primaryClass = {astro-ph},
       adsurl = {https://ui.adsabs.harvard.edu/abs/1995PASP..107..803U}
}

@ARTICLE{2019ApJ...874L..29R,
       author = {{Rodrigues}, Xavier and {Gao}, Shan and {Fedynitch}, Anatoli and {Palladino}, Andrea and {Winter}, Walter},
        title = "{Leptohadronic Blazar Models Applied to the 2014-2015 Flare of TXS 0506+056}",
      journal = {\apjl},
         year = 2019,
        month = apr,
       volume = {874},
       number = {2},
          eid = {L29},
        pages = {L29},
          doi = {10.3847/2041-8213/ab1267},
archivePrefix = {arXiv},
       eprint = {1812.05939},
 primaryClass = {astro-ph.HE},
       adsurl = {https://ui.adsabs.harvard.edu/abs/2019ApJ...874L..29R}
}

@ARTICLE{2019NatAs...3...88G,
       author = {{Gao}, Shan and {Fedynitch}, Anatoli and {Winter}, Walter and {Pohl}, Martin},
        title = "{Modelling the coincident observation of a high-energy neutrino and a bright blazar flare}",
      journal = {Nature Astronomy},
         year = 2019,
        month = jan,
       volume = {3},
        pages = {88-92},
          doi = {10.1038/s41550-018-0610-1},
archivePrefix = {arXiv},
       eprint = {1807.04275},
 primaryClass = {astro-ph.HE},
       adsurl = {https://ui.adsabs.harvard.edu/abs/2019NatAs...3...88G}
}

@ARTICLE{2019MNRAS.483L..12C,
       author = {{Cerruti}, M. and {Zech}, A. and {Boisson}, C. and {Emery}, G. and {Inoue}, S. and {Lenain}, J.-P.},
        title = "{Leptohadronic single-zone models for the electromagnetic and neutrino emission of TXS 0506+056}",
      journal = {\mnras},
         year = 2019,
        month = feb,
       volume = {483},
       number = {1},
        pages = {L12-L16},
          doi = {10.1093/mnrasl/sly210},
archivePrefix = {arXiv},
       eprint = {1807.04335},
 primaryClass = {astro-ph.HE},
       adsurl = {https://ui.adsabs.harvard.edu/abs/2019MNRAS.483L..12C}
}

@ARTICLE{2018ApJ...864...84K,
       author = {{Keivani}, A. and {Murase}, K. and {Petropoulou}, M. and {Fox}, D.~B. and {Cenko}, S.~B. and {Chaty}, S. and {Coleiro}, A. and {DeLaunay}, J.~J. and {Dimitrakoudis}, S. and {Evans}, P.~A. and {Kennea}, J.~A. and {Marshall}, F.~E. and {Mastichiadis}, A. and {Osborne}, J.~P. and {Santander}, M. and {Tohuvavohu}, A. and {Turley}, C.~F.},
        title = "{A Multimessenger Picture of the Flaring Blazar TXS 0506+056: Implications for High-energy Neutrino Emission and Cosmic-Ray Acceleration}",
      journal = {\apj},
         year = 2018,
        month = sep,
       volume = {864},
       number = {1},
          eid = {84},
        pages = {84},
          doi = {10.3847/1538-4357/aad59a},
archivePrefix = {arXiv},
       eprint = {1807.04537},
 primaryClass = {astro-ph.HE},
       adsurl = {https://ui.adsabs.harvard.edu/abs/2018ApJ...864...84K}
}

@ARTICLE{2022ApJ...927..197A,
       author = {{Acciari}, V.~A. and {Aniello}, T. and {Ansoldi}, S. and {Antonelli}, L.~A. and {Arbet Engels}, A. and {Artero}, M. and {Asano}, K. and {Baack}, D. and {Babi{\'c}}, A. and {Baquero}, A. and {Barres de Almeida}, U. and {Barrio}, J.~A. and {Batkovi{\'c}}, I. and {Becerra Gonz{\'a}lez}, J. and {Bednarek}, W. and {Bernardini}, E. and {Bernardos}, M. and {Berti}, A. and {Besenrieder}, J. and {Bhattacharyya}, W. and {Bigongiari}, C. and {Biland}, A. and {Blanch}, O. and {B{\"o}kenkamp}, H. and {Bonnoli}, G. and {Bo{\v{s}}njak}, {\v{Z}}. and {Busetto}, G. and {Carosi}, R. and {Ceribella}, G. and {Cerruti}, M. and {Chai}, Y. and {Chilingarian}, A. and {Cikota}, S. and {Colombo}, E. and {Contreras}, J.~L. and {Cortina}, J. and {Covino}, S. and {D'Amico}, G. and {D'Elia}, V. and {Vela}, P. Da and {Dazzi}, F. and {De Angelis}, A. and {De Lotto}, B. and {Del Popolo}, A. and {Delfino}, M. and {Delgado}, J. and {Mendez}, C. Delgado and {Depaoli}, D. and {Di Pierro}, F. and {Di Venere}, L. and {Do Souto Espi{\~n}eira}, E. and {Dominis Prester}, D. and {Donini}, A. and {Dorner}, D. and {Doro}, M. and {Elsaesser}, D. and {Fallah Ramazani}, V. and {Fari{\~n}a}, L. and {Fattorini}, A. and {Font}, L. and {Fruck}, C. and {Fukami}, S. and {Fukazawa}, Y. and {Garc{\'\i}a L{\'o}pez}, R.~J. and {Garczarczyk}, M. and {Gasparyan}, S. and {Gaug}, M. and {Giglietto}, N. and {Giordano}, F. and {Gliwny}, P. and {Godinovi{\'c}}, N. and {Green}, J.~G. and {Green}, D. and {Hadasch}, D. and {Hahn}, A. and {Hassan}, T. and {Heckmann}, L. and {Herrera}, J. and {Hoang}, J. and {Hrupec}, D. and {H{\"u}tten}, M. and {Inada}, T. and {Iotov}, R. and {Ishio}, K. and {Iwamura}, Y. and {Jim{\'e}nez Mart{\'\i}nez}, I. and {Jormanainen}, J. and {Jouvin}, L. and {Kerszberg}, D. and {Kobayashi}, Y. and {Kubo}, H. and {Kushida}, J. and {Lamastra}, A. and {Lelas}, D. and {Leone}, F. and {Lindfors}, E. and {Linhoff}, L. and {Lombardi}, S. and {Longo}, F. and {L{\'o}pez-Coto}, R. and {L{\'o}pez-Moya}, M. and {L{\'o}pez-Oramas}, A. and {Loporchio}, S. and {Machado de Oliveira Fraga}, B. and {Maggio}, C. and {Majumdar}, P. and {Makariev}, M. and {Mallamaci}, M. and {Maneva}, G. and {Manganaro}, M. and {Mannheim}, K. and {Mariotti}, M. and {Mart{\'\i}nez}, M. and {Mas Aguilar}, A. and {Mazin}, D. and {Menchiari}, S. and {Mender}, S. and {Mi{\'c}anovi{\'c}}, S. and {Miceli}, D. and {Miener}, T. and {Miranda}, J.~M. and {Mirzoyan}, R. and {Molina}, E. and {Moralejo}, A. and {Morcuende}, D. and {Moreno}, V. and {Moretti}, E. and {Nakamori}, T. and {Nava}, L. and {Neustroev}, V. and {Nievas Rosillo}, M. and {Nigro}, C. and {Nilsson}, K. and {Nishijima}, K. and {Noda}, K. and {Nozaki}, S. and {Ohtani}, Y. and {Oka}, T. and {Otero-Santos}, J. and {Paiano}, S. and {Palatiello}, M. and {Paneque}, D. and {Paoletti}, R. and {Paredes}, J.~M. and {Pavleti{\'c}}, L. and {Pe{\~n}il}, P. and {Persic}, M. and {Pihet}, M. and {Prada Moroni}, P.~G. and {Prandini}, E. and {Priyadarshi}, C. and {Puljak}, I. and {Rhode}, W. and {Rib{\'o}}, M. and {Rico}, J. and {Righi}, C. and {Rugliancich}, A. and {Sahakyan}, N. and {Saito}, T. and {Sakurai}, S. and {Satalecka}, K. and {Saturni}, F.~G. and {Schleicher}, B. and {Schmidt}, K. and {Schmuckermaier}, F. and {Schweizer}, T. and {Sitarek}, J. and {{\v{S}}nidari{\'c}}, I. and {Sobczynska}, D. and {Spolon}, A. and {Stamerra}, A. and {Stri{\v{s}}kovi{\'c}}, J. and {Strom}, D. and {Strzys}, M. and {Suda}, Y. and {Suri{\'c}}, T. and {Takahashi}, M. and {Takeishi}, R. and {Tavecchio}, F. and {Temnikov}, P. and {Terzi{\'c}}, T. and {Teshima}, M. and {Tosti}, L. and {Truzzi}, S. and {Tutone}, A. and {Ubach}, S. and {van Scherpenberg}, J. and {Vanzo}, G. and {Vazquez Acosta}, M. and {Ventura}, S. and {Verguilov}, V. and {Viale}, I. and {Vigorito}, C.~F. and {Vitale}, V. and {Vovk}, I. and {Will}, M. and {Wunderlich}, C. and {Yamamoto}, T. and {Zari{\'c}}, D. and {Hodges}, M.},
        title = "{Investigating the Blazar TXS 0506+056 through Sharp Multiwavelength Eyes During 2017-2019}",
        journal = {\apj},
         year = 2022,
        month = mar,
       volume = {927},
       number = {2},
          eid = {197},
        pages = {197},
          doi = {10.3847/1538-4357/ac531d},
archivePrefix = {arXiv},
       eprint = {2202.02600},
 primaryClass = {astro-ph.HE},
       adsurl = {https://ui.adsabs.harvard.edu/abs/2022ApJ...927..197A}
}

@ARTICLE{2020ApJ...891..115P,
       author = {{Petropoulou}, Maria and {Murase}, Kohta and {Santander}, Marcos and {Buson}, Sara and {Tohuvavohu}, Aaron and {Kawamuro}, Taiki and {Vasilopoulos}, Georgios and {Negoro}, Hiroshi and {Ueda}, Yoshihiro and {Siegel}, Michael H. and {Keivani}, Azadeh and {Kawai}, Nobuyuki and {Mastichiadis}, Apostolos and {Dimitrakoudis}, Stavros},
        title = "{Multi-epoch Modeling of TXS 0506+056 and Implications for Long-term High-energy Neutrino Emission}",
      journal = {\apj},
         year = 2020,
        month = mar,
       volume = {891},
       number = {2},
          eid = {115},
        pages = {115},
          doi = {10.3847/1538-4357/ab76d0},
archivePrefix = {arXiv},
       eprint = {1911.04010},
 primaryClass = {astro-ph.HE},
       adsurl = {https://ui.adsabs.harvard.edu/abs/2020ApJ...891..115P}
}

@ARTICLE{2024ApJS..275....4K,
       author = {{Klinger}, Marc and {Rudolph}, Annika and {Rodrigues}, Xavier and {Yuan}, Chengchao and {Fichet de Clairfontaine}, Ga{\"e}tan and {Fedynitch}, Anatoli and {Winter}, Walter and {Pohl}, Martin and {Gao}, Shan},
        title = "{AM$^{3}$: An Open-source Tool for Time-dependent Lepto-hadronic Modeling of Astrophysical Sources}",
      journal = {\apjs},
         year = 2024,
        month = nov,
       volume = {275},
       number = {1},
          eid = {4},
        pages = {4},
          doi = {10.3847/1538-4365/ad725c},
archivePrefix = {arXiv},
       eprint = {2312.13371},
 primaryClass = {astro-ph.HE},
       adsurl = {https://ui.adsabs.harvard.edu/abs/2024ApJS..275....4K}
}

@ARTICLE{2018MNRAS.480..192P,
       author = {{Padovani}, P. and {Giommi}, P. and {Resconi}, E. and {Glauch}, T. and {Arsioli}, B. and {Sahakyan}, N. and {Huber}, M.},
        title = "{Dissecting the region around IceCube-170922A: the blazar TXS 0506+056 as the first cosmic neutrino source}",
      journal = {\mnras},
         year = 2018,
        month = oct,
       volume = {480},
       number = {1},
        pages = {192-203},
          doi = {10.1093/mnras/sty1852},
archivePrefix = {arXiv},
       eprint = {1807.04461},
 primaryClass = {astro-ph.HE},
       adsurl = {https://ui.adsabs.harvard.edu/abs/2018MNRAS.480..192P}
}

@ARTICLE{2025Univ...11..204M,
       author = {{Miao}, Xianglin and {Jiang}, Yunguo},
        title = "{Investigating the Variation and Periodicity of TXS 0506+056}",
      journal = {Universe},
         year = 2025,
        month = jun,
       volume = {11},
       number = {7},
          eid = {204},
        pages = {204},
          doi = {10.3390/universe11070204},
       adsurl = {https://ui.adsabs.harvard.edu/abs/2025Univ...11..204M}
}

@ARTICLE{2012AJ....143..108W,
       author = {{Wu}, Jianghua and {B{\"o}ttcher}, Markus and {Zhou}, Xu and {He}, Xiangtao and {Ma}, Jun and {Jiang}, Zhaoji},
        title = "{Simultaneous B'V'R' Monitoring of BL Lacertae Object S5 0716+714 and Detection of Inter-band Time Delay}",
      journal = {\aj},
         year = 2012,
        month = may,
       volume = {143},
       number = {5},
          eid = {108},
        pages = {108},
          doi = {10.1088/0004-6256/143/5/108},
archivePrefix = {arXiv},
       eprint = {1202.3226},
 primaryClass = {astro-ph.HE},
       adsurl = {https://ui.adsabs.harvard.edu/abs/2012AJ....143..108W}
}

@ARTICLE{2022ApJ...933..224F,
       author = {{Fang}, Yue and {Chen}, Qihang and {Zhang}, Yan and {Wu}, Jianghua},
        title = "{Multiwavelength Variation Phenomena of PKS 0735+178 on Diverse Timescales}",
      journal = {\apj},
         year = 2022,
        month = jul,
       volume = {933},
       number = {2},
          eid = {224},
        pages = {224},
          doi = {10.3847/1538-4357/ac7647},
archivePrefix = {arXiv},
       eprint = {2206.03296},
 primaryClass = {astro-ph.GA},
       adsurl = {https://ui.adsabs.harvard.edu/abs/2022ApJ...933..224F}
}

@ARTICLE{2018MNRAS.478.3513Z,
       author = {{Zhang}, Xiaoyuan and {Wu}, Jianghua and {Meng}, Nankun},
        title = "{Intra-day optical multi-band quasi-simultaneous observation of BL Lacertae object S5 0716+714 from 2013 to 2016}",
      journal = {\mnras},
         year = 2018,
        month = aug,
       volume = {478},
       number = {3},
        pages = {3513-3524},
          doi = {10.1093/mnras/sty1468},
archivePrefix = {arXiv},
       eprint = {1806.08148},
 primaryClass = {astro-ph.GA},
       adsurl = {https://ui.adsabs.harvard.edu/abs/2018MNRAS.478.3513Z}
}

@ARTICLE{2018ApJ...862..123M,
       author = {{Mudd}, D. and {Martini}, P. and {Zu}, Y. and {Kochanek}, C. and {Peterson}, B.~M. and {Kessler}, R. and {Davis}, T.~M. and {Hoormann}, J.~K. and {King}, A. and {Lidman}, C. and {Sommer}, N.~E. and {Tucker}, B.~E. and {Asorey}, J. and {Hinton}, S. and {Glazebrook}, K. and {Kuehn}, K. and {Lewis}, G. and {Macaulay}, E. and {Moeller}, A. and {O'Neill}, C. and {Zhang}, B. and {Abbott}, T.~M.~C. and {Abdalla}, F.~B. and {Allam}, S. and {Banerji}, M. and {Benoit-L{\'e}vy}, A. and {Bertin}, E. and {Brooks}, D. and {Carnero Rosell}, A. and {Carollo}, D. and {Carrasco Kind}, M. and {Carretero}, J. and {Cunha}, C.~E. and {D'Andrea}, C.~B. and {da Costa}, L.~N. and {Davis}, C. and {Desai}, S. and {Doel}, P. and {Fosalba}, P. and {Garc{\'\i}a-Bellido}, J. and {Gaztanaga}, E. and {Gerdes}, D.~W. and {Gruen}, D. and {Gruendl}, R.~A. and {Gschwend}, J. and {Gutierrez}, G. and {Hartley}, W.~G. and {Honscheid}, K. and {James}, D.~J. and {Kuhlmann}, S. and {Kuropatkin}, N. and {Lima}, M. and {Maia}, M.~A.~G. and {Marshall}, J.~L. and {McMahon}, R.~G. and {Menanteau}, F. and {Miquel}, R. and {Plazas}, A.~A. and {Romer}, A.~K. and {Sanchez}, E. and {Schindler}, R. and {Schubnell}, M. and {Smith}, M. and {Smith}, R.~C. and {Soares-Santos}, M. and {Sobreira}, F. and {Suchyta}, E. and {Swanson}, M.~E.~C. and {Tarle}, G. and {Thomas}, D. and {Tucker}, D.~L. and {Walker}, A.~R. and {DES Collaboration}},
        title = "{Quasar Accretion Disk Sizes from Continuum Reverberation Mapping from the Dark Energy Survey}",
      journal = {\apj},
         year = 2018,
        month = aug,
       volume = {862},
       number = {2},
          eid = {123},
        pages = {123},
          doi = {10.3847/1538-4357/aac9bb},
archivePrefix = {arXiv},
       eprint = {1711.11588},
 primaryClass = {astro-ph.GA},
       adsurl = {https://ui.adsabs.harvard.edu/abs/2018ApJ...862..123M}
}

@ARTICLE{2013arXiv1302.1508A,
       author = {{Alexander}, Tal},
        title = "{Improved AGN light curve analysis with the z-transformed discrete correlation function}",
      journal = {arXiv e-prints},
         year = 2013,
        month = feb,
          eid = {arXiv:1302.1508},
        pages = {arXiv:1302.1508},
          doi = {10.48550/arXiv.1302.1508},
archivePrefix = {arXiv},
       eprint = {1302.1508},
 primaryClass = {astro-ph.IM},
       adsurl = {https://ui.adsabs.harvard.edu/abs/2013arXiv1302.1508A}
}

@INPROCEEDINGS{1997ASSL..218..163A,
       author = {{Alexander}, Tal},
        title = "{Is AGN Variability Correlated with Other AGN Properties? ZDCF Analysis of Small Samples of Sparse Light Curves}",
    booktitle = {Astronomical Time Series},
         year = 1997,
       editor = {{Maoz}, D. and {Sternberg}, A. and {Leibowitz}, E.~M.},
       series = {Astrophysics and Space Science Library},
       volume = {218},
        month = jan,
        pages = {163},
          doi = {10.1007/978-94-015-8941-3_14},
       adsurl = {https://ui.adsabs.harvard.edu/abs/1997ASSL..218..163A}
}
\bibliographystyle{aasjournalv7}

\end{document}